\documentclass [12pt,preprint]{aastex}
\usepackage{epsfig}

\begin{document}
\bibliographystyle{plainnat}

\newcommand{\xmm}{{\small \it XMM-Newton}}
\newcommand{\swift}{{\small \it Swift}}
\newcommand{\rosat}{{\small \it Rosat}}
\newcommand{\gsim}{\hbox{\rlap{$^>$}$_\sim$}}
\newcommand{\lsim}{\hbox{\rlap{$^<$}$_\sim$}}
\newcommand \x {X-ray}
\newcommand{\xspec}{{\small \it Xspec}}
\newcommand{\pn}{{\small EPIC~PN}}

\title{Can the soft X-ray opacity towards high redshift sources probe
the missing baryons?}

\author{Ehud Behar\altaffilmark{1},  Shlomo Dado\altaffilmark{1},  
Arnon Dar\altaffilmark{1} and Ari Laor\altaffilmark{1}}

\altaffiltext{1}{Physics Department, Technion, Haifa 32000, Israel.
behar, dado, arnon, laor@physics.technion.ac.il}

\begin{abstract}

Observations with the \swift\ satellite of \x\ afterglows of more than a hundred gamma ray bursts (GRBs) with known redshift reveal ubiquitous soft X-ray absorption. 
The directly measured optical depth $\tau$ at a given observed energy is found to be constant on average at redshift $z > 2$, i.e., $\langle \tau (0.5~\mathrm{keV}) \rangle_{z > 2}\, = 0.40\pm 0.02$.
Such an asymptotic optical depth is expected if the foreground diffuse intergalactic medium (IGM) dominates the absorption effect, 
and if the metallicity of the diffuse IGM reaches $\sim 0.2 - 0.4$ solar at $z = 0$.
To further test the IGM absorption hypothesis, we analyze the 12 highest S/N ($> 5000$ photon) $z > 2$ quasar spectra from the \xmm\ archive, which are all extremely radio loud (RLQs).
The quasar optical depths are found to be 
consistent with the mean GRB value. 
The four lowest-$z$ quasars ($2 < z < 2.5$), however, do not show significant absorption.
The best \x\ spectra of radio-quiet quasars (RQQs) at $z > 2$ provide only upper limits to the absorption, which are still consistent with the RLQs, albeit with much lower S/N ($\lsim 1000$ photons at $z \approx 4$).
Lack of quasar absorption poses a challenge to the smooth IGM interpretation,
and could allude to the opacity being rather due to the jets in RLQs and GRBs.
However, the jet absorbing column would need to appear in RLQs only at $z~\gsim\ 2.5$, and in GRBs to strongly increase with $z$ in order to produce the observed tendency to a constant mean $\tau$.
High \x\ spectral resolution can differentiate between an absorber intrinsic to the source that produces discernible spectral lines, and the diffuse IGM that produces significant absorption, but no discrete features.

\end{abstract}


\maketitle

\section{Introduction} 
\label{into}

\swift\ \x\ spectra of gamma ray bursts (GRB) afterglows reveal prevalent soft X-ray
absorption, which is commonly assumed to originate in the host galaxy.
The standard \x\ absorption measurement technique probes metal absorption, but quotes equivalent hydrogen column densities $N_H$ by assuming a neutral absorber at the host ($z$), and with solar abundances.
Under these assumptions, $N_H$ towards GRBs shows a strong correlation with the host redshift $z$.
The typical absorbing column rises from $N_{\rm H}\sim 10^{21}$~cm$^{-2}$ 
at $z<1$ up to  $\approx 10^{23}$~cm$^{-2}$ at the highest observed redshifts
\citep{Jakobsson2006, Campana2006, Watson2007, Campana2010, Rau2010}.
Damped Lyman-$\alpha$ absorption is also seen in some GRB afterglows.
Unlike the \x\ absorber, the redshift of the Lyman-$\alpha$ absorber is well constrained.
The implied column is usually still well below the X-ray derived column
\citep[see, e.g.,][]{Watson2007}.
If the abundances in the \x\ absorber are sub-solar, the \x\ derived $N_H$ values are even higher and the discrepancy with the Lyman-$\alpha$ column grows accordingly.
Although an appreciable ionization range in a single medium can possibly account for such a discrepancy \citep[recently,][]{schady10},
there is also the possibility that the \x\ and Lyman-$\alpha$ absorbers are physically distinct. 

High-$z$ {\it quasars} also commonly reveal soft X-ray
absorption \citep[e.g.,][]{Fabian2001, Worsley2004a, Worsley2004b, Page2005, Yuan2005, Grupe2006, sambruna07}.
The implied absoring column, if intrinsic, is also of the order of $10^{23}$~cm$^{-2}$.
In the case of quasars as well, metal and hydrogen
line absorption in the UV imply significantly lower $N_H$ columns.
In the case of low-luminosity active galaxies, partially ionized outflows are known to have more \x\ column with only a trace 
of UV absorbing ions \citep{crenshaw03},
a discrepancy that again can be partially reconciled with an ionization correction. 
The X-ray absorption profile in three different quasars at $z\sim 4.3-4.7$ was noted by \citet{Yuan2005} to be remarkably similar, while ionized quasar outflows are not necessarily expected to be so uniform.
Indeed, partially ionized outflows have not been directly identified in such luminous quasars,
and thus it is plausible that the \x\ and UV absorbers towards high-$z$ quasars are also physically distinct.

Intrigued by the aforementioned puzzles associated with soft \x\ absorption of high-$z$ sources, we wish to explore  the similarities of \x\ absorption of high-z GRBs and quasars and their possible origin. 
In Sec.~\ref{GRB} we present the \x\ opacities towards GRB afterglows from the \swift\ sample \citep{Evans2009, Campana2010}, but now without assuming neither a redshift, nor an ionization, nor a metallicity for the \x\ absorber.
We then discuss in Sec.~\ref{IGM} the viability of \x\ absorption by the diffuse intergalactic medium (IGM).
In Sec.~\ref{quasars}, we present a comparison sample of the highest signal to noise ratio (S/N) $z > 2$ quasar spectra drawn from the \xmm\ archive.
Realizing that our selection criterion based on the number of detected photons allows only radio loud quasars into the sample, in Sec.~\ref{RQQ} we also present the best S/N high-z radio quiet quasars from the works of \citet{shemmer08,shemmer06,shemmer05}.
In Sec.~\ref{concl} we conclude and compare the expected absorption signature in the \x\ spectra of high-$z$ sources of a diffuse IGM versus that of jets and propose future tests that can reveal whether the observed absorption is indeed due to the IGM or intrinsic to the sources.

\section{Observed-frame opacity of GRBs}
\label{GRB}

Soft \x\ absorption below 1 keV is mostly due to photo-ionization 
of heavy elements (e.g., C, N, O, Fe) and has been measured with the
\swift\ \x\ Telescope (XRT) in over a hundred GRBs with an identified redshift \citep{Gehrels2004, Evans2009, Campana2010}. 
It is preferable to use late photon arrival times when spectral variability is minimal in order to obtain the most reliable absorption measurements.
In this work, we therefore use only the photon counting (PC) mode data available from the \swift /GRB spectrum repository \citep[][\footnote{http://www.swift.ac.uk/xrt\_spectra/}]{Evans2009}.
This choice of data is not as sophisticated as the careful time cuts of \citet{Campana2010}, but on the other hand, it provides a uniformly reduced sample.
We include all of the bursts up to 2010 July.
We use only the 144 that have an identified redshift, out of a total of 520 GRBs.
The sample greatly varies in S/N, but in order not to introduce any biases, we do not exclude any objects from the analysis.
Out of the sample of 144 GRBs, absorption at the 90\% confidence level is measured for 113, while the other 31 have only upper limits.

\begin{figure}
\vglue13.0cm
\includegraphics[width=\columnwidth]{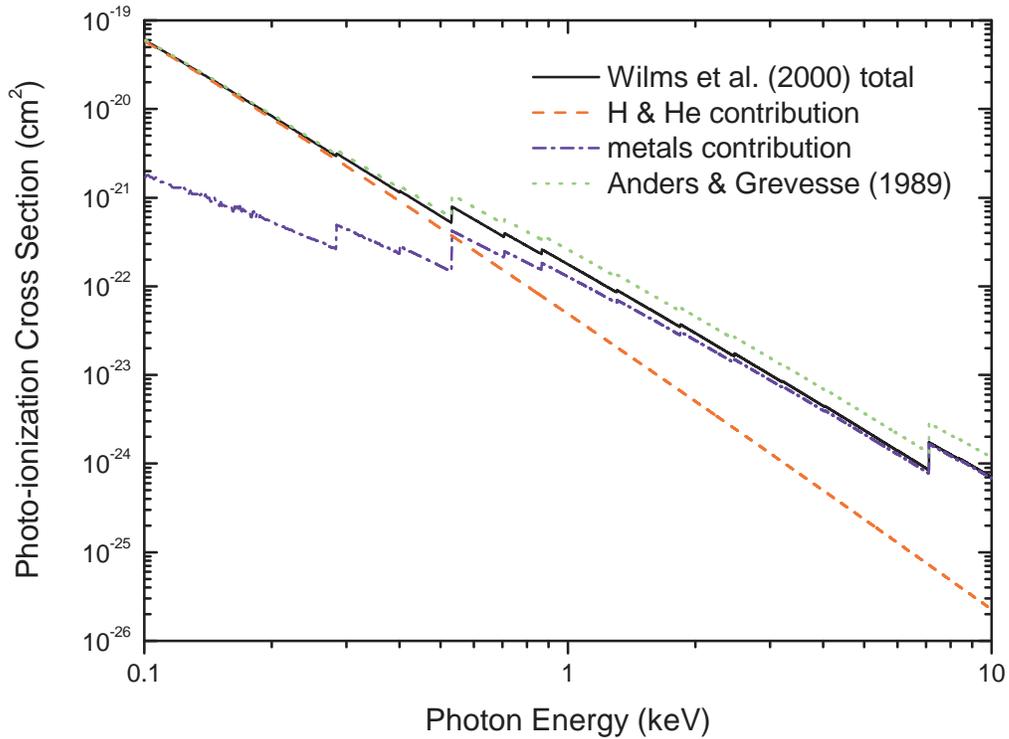} 
\vglue-5.5cm
\caption{Photo-ionization cross section per H atom extracted from the \xspec\ {\it phabs} model using the \citet{Wilms2000} abundances preferred in this work (solid line), and separated into its H and He contribution (dashed) and metal contribution (dash-dot). The cross section obtained with the \citet{Anders1989} abundances (dotted line) are shown for comparison.}
\label{fig:cs}
\end{figure}

The column density $N_H(z)$ in the repository is conveniently given as an equivalent hydrogen column density assuming a neutral solar-abundance absorber at the host redshift.
The strong correlation of these absorbing columns with $z$ was noted by \citet{Campana2010}. 
Here, we wish to  carefully study the observed \x\ absorption effect,
 relaxing these assumptions as much as possible.
We thus convert the quoted column into an optical depth at the observed energy $E$ using 

\begin{equation} 
\tau ^*(E)=\sigma[(1+z)E]N_H(z) 
\label{tau*} 
\end{equation} 

\noindent where $\sigma([1+z]E)$ is the total photo-ionization cross section 
per hydrogen atom at a photon energy of $[1+z]E$ and for solar metallicity gas.
In Eq.~\ref{tau*}, we use the exact same cross section used by the repository team,
namely that provided by the {\it phabs} model in \xspec \footnote{http://heasarc.gsfc.nasa.gov/docs/xanadu/xspec/} for a neutral solar-composition \citep{Anders1989} absorber.
Importantly, the intrinsic column densities provided in the repository are those in excess of the nominal local Galactic column  $N^\mathrm{Gal} _H$ based on HI 21~cm measurements \citep{kalberia05}, which is also given there.
However, we note that the use of the solar abundances of \citet{Anders1989} for the Galactic column leads to an overestimate of the local \x\ absorption effect and, thus, to an underestimate of the residual (GRB)  absorption, as these abundances are somewhat high compared to more up to date estimates of interstellar abundances \citep{Wilms2000}.
At 0.5~keV, e.g., the photo-ionization cross section per H atom based on \citet{Anders1989} abundances  would be 7.14$\times 10^{-22}$ cm$^2$ compared to the preferred value of 6.22$\times 10^{-22}$ cm$^2$ obtained with the \citet{Wilms2000} abundances.
This difference is demonstrated in Fig.~\ref{fig:cs}.
We therefore introduce a small correction (increase) to the optical depth in Eq.~\ref{tau*} as follows:

\begin{equation} 
\tau(E)=\tau ^*(E) + N^\mathrm{Gal} _H[\sigma ^\mathrm{AG} (E) - \sigma ^\mathrm{W} (E) ]
\label{tau} 
\end{equation} 

\noindent where $\sigma ^\mathrm{AG} (E) - \sigma ^\mathrm{W} (E)$ is the (positive) difference in the inferred cross section deduced from using the \citet[][AG]{Anders1989} and \citet[][W]{Wilms2000} abundances.  On average, this correction to $\tau$ is $\sim 10\%$, but can reach 50\% for high Galactic column GRBs. 

\begin{figure}[pt!]
{\includegraphics[width=13cm, angle=-90]{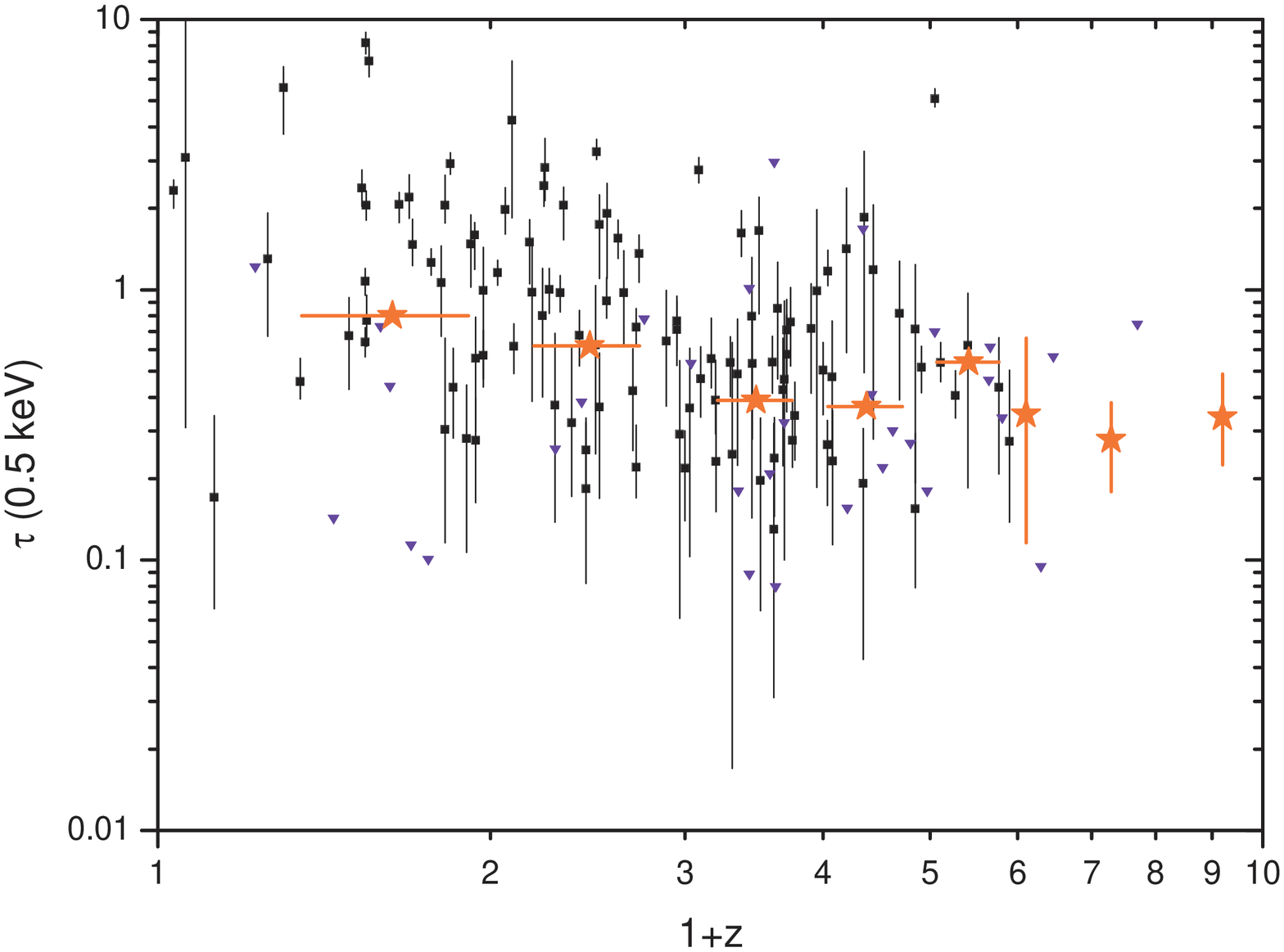}}
\caption{Optical depth $\tau$ at 0.5 keV as a function of redshift for the \swift /XRT PC-mean GRB sample.
Black squares are detections and red stars represent error-weighted $\langle \tau \rangle$ values averaged over $\Delta z = 1$ bins; 
Note the striking constancy of $\langle \tau \rangle$ with $z$ and its tendency towards the value of $\sim0.4$ for $z > 2$.  
Blue triangles represent +90\% confidence (upper) limits. }
\label{fig:tau} 
\end{figure}

In Fig.~\ref{fig:tau} we present the PC-mean results for the optical depth $\tau (0.5~\mathrm{keV})$ at the observed energy of 0.5~keV, where photo-electric absorption is significant and the instrument response is sufficiently high.
The errors on $\tau$ correspond to the 90\% confidence errors quoted in the repository.
For bursts whose absorption measurements are consistent with zero to within these errors, we plot only the +90\% confidence (upper) limit.
We stress that the optical depth is the directly measurable model-independent quantity, unlike column density.
The scatter in $\tau (0.5\, \mathrm{keV})$ in Fig.~\ref{fig:tau} is substantial, but there is no evolution with $z$, especially not at high $z$.
Indeed, the error-weighted mean optical depth in $\Delta z = 1$ bins is also shown in the figure, and its constancy with $z$ is strikingly tight.
For $z > 2$ bursts, the error-weighted mean optical depth is $\langle \tau (0.5~\mathrm{keV})\rangle _{z>2}\,= 0.40 \pm 0.02$.
The upper limits can not be included in an error-weighted mean, but 
a simple mean with no weighing yields $\langle \tau (0.5~\mathrm{keV}) \rangle _{z>2}\, =0.7 \pm 0.1$ with the upper limits, and $\langle \tau (0.5~\mathrm{keV})\rangle _{z>2}\,= 0.55 \pm 0.09$ without them;
Quoted errors are standard errors of the mean.
Using the late-time PC spectra from the repository instead of the mean PC data increases the unweighed mean by $\approx 0.1$.
This could be a result of the 2--3\% steeper spectral slopes in the late-time spectra. 
On the other hand, excluding GRB\,060202, a rare outlier with $\tau(0.5\, \mathrm{keV}) > 5$, which also has a very steep spectral slope of $\Gamma = 2.7$ and is not fitted by the model very well, reduces the unweighed mean by $\approx 0.1$.
All of these give a general idea of the systematic uncertainties that can be expected from the \x\ absorption measurements of the transient GRB afterglows.
Given the wide range in the quality of the measurements, their detection limits, and the fact that including upper limits tends to {\it increase} the mean (i.e., largely inadequate measurements rather than low absorption), we prefer the error weighted mean of $\langle \tau(0.5\, \mathrm{keV})\rangle _{z>2}\, = 0.40 \pm 0.02$, keeping in mind, however, the large ($\sim 0.2$) systematic errors.

\begin{figure}[pt!]
\begin{center}
\vglue0.0cm
{\includegraphics[angle=-90,width=5.3cm]{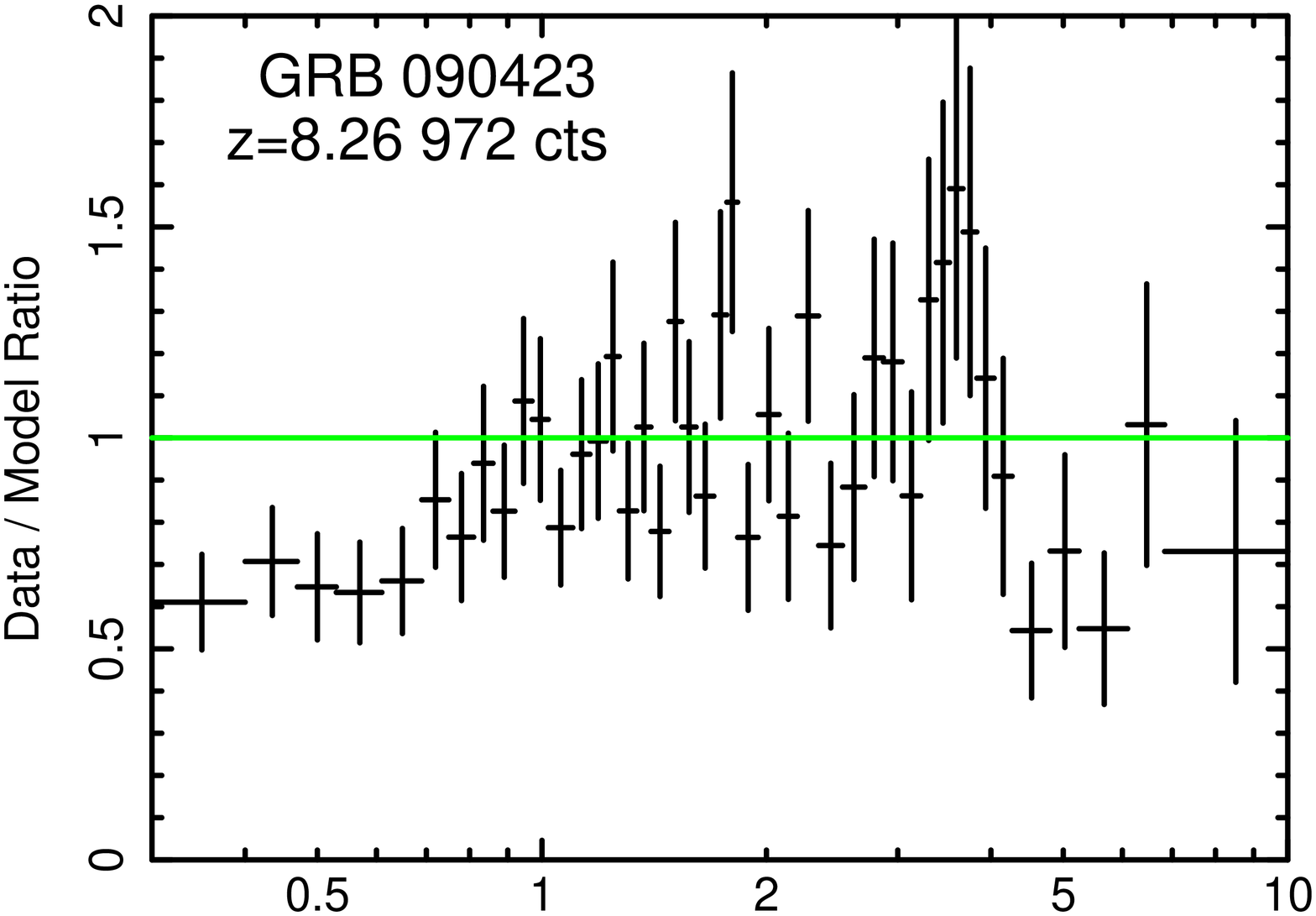}}
{\includegraphics[angle=-90,width=5cm]{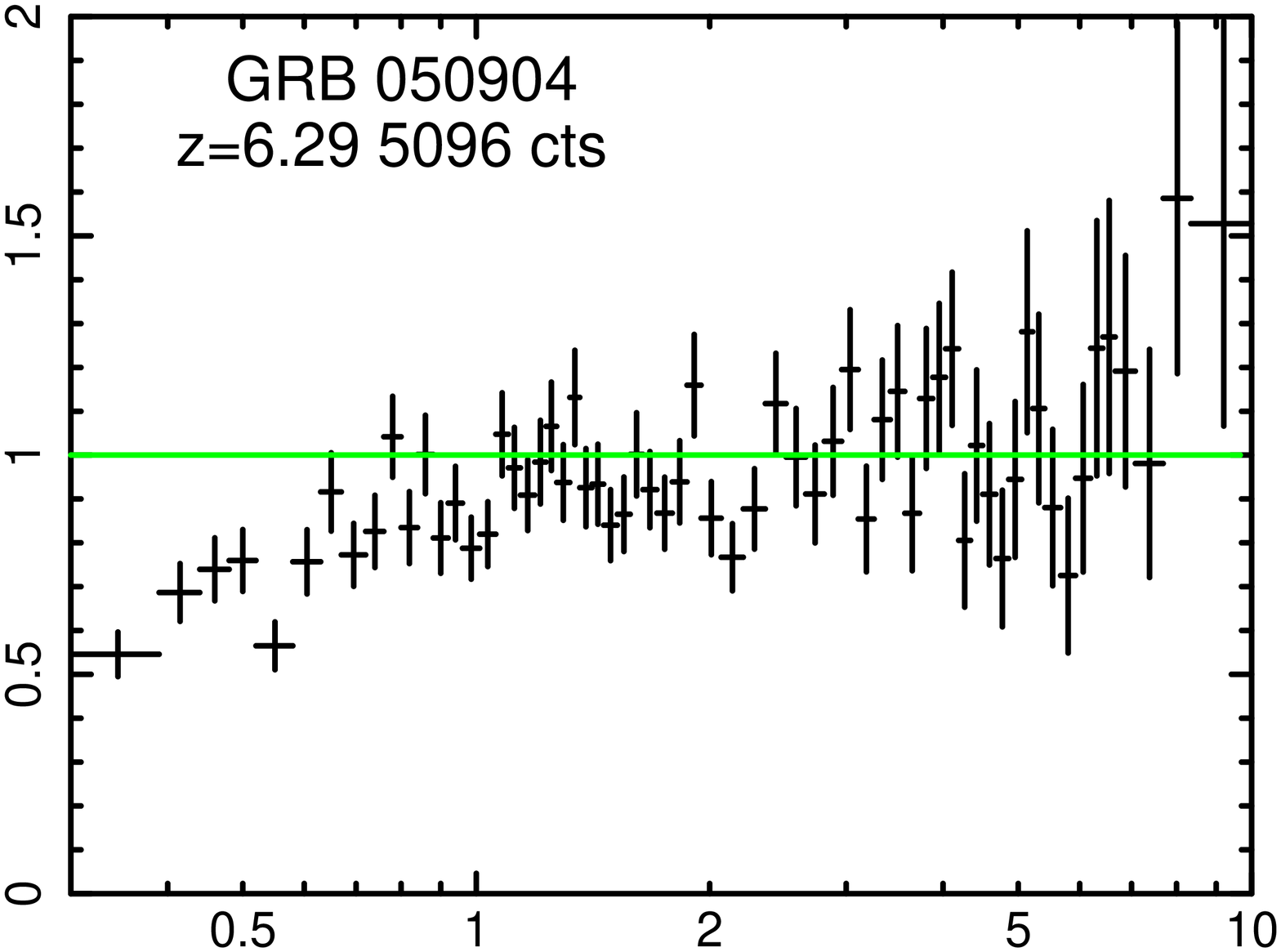}}
{\includegraphics[angle=-90,width=5.cm]{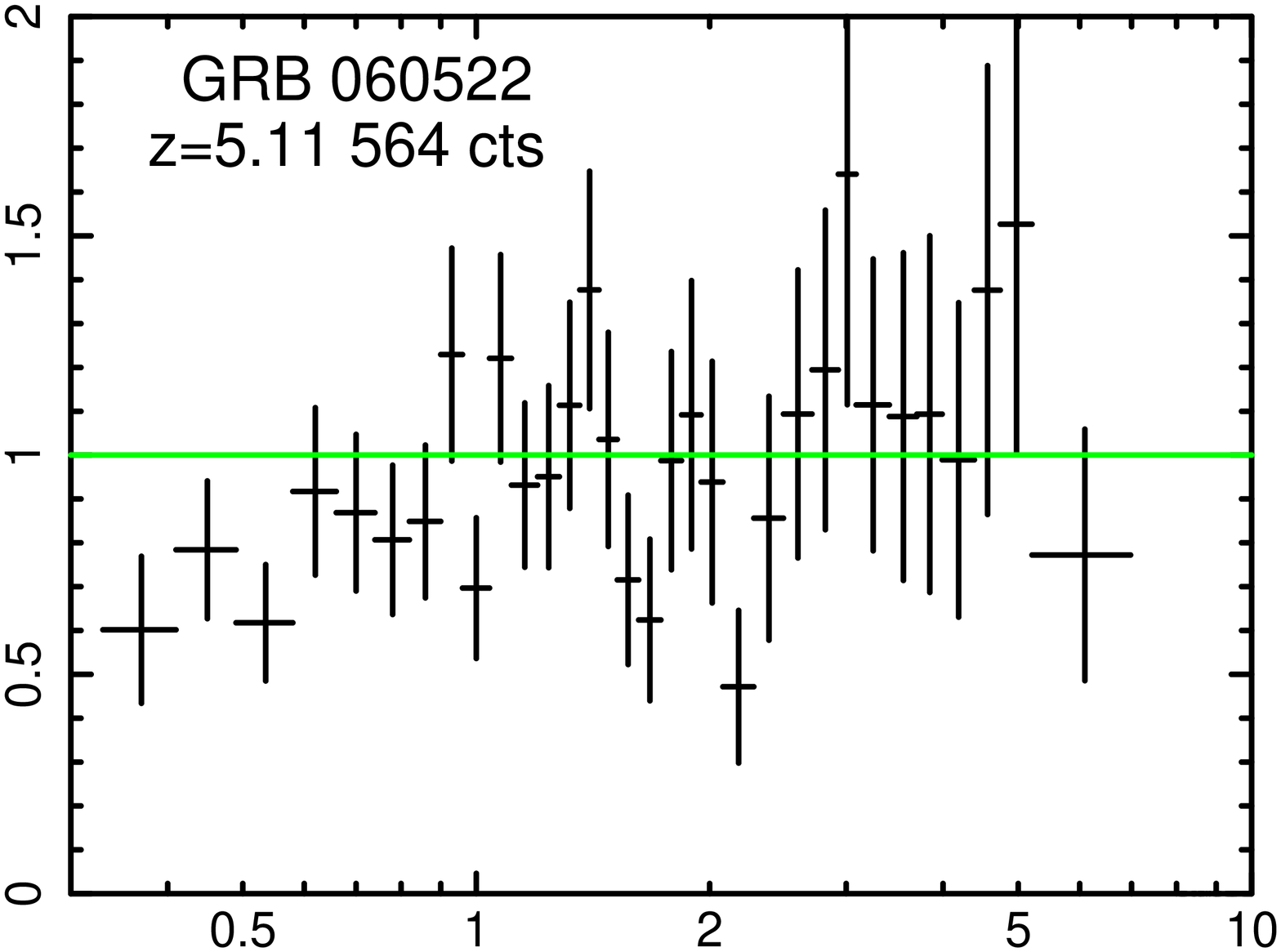}}
{\includegraphics[angle=-90,width=5.3cm]{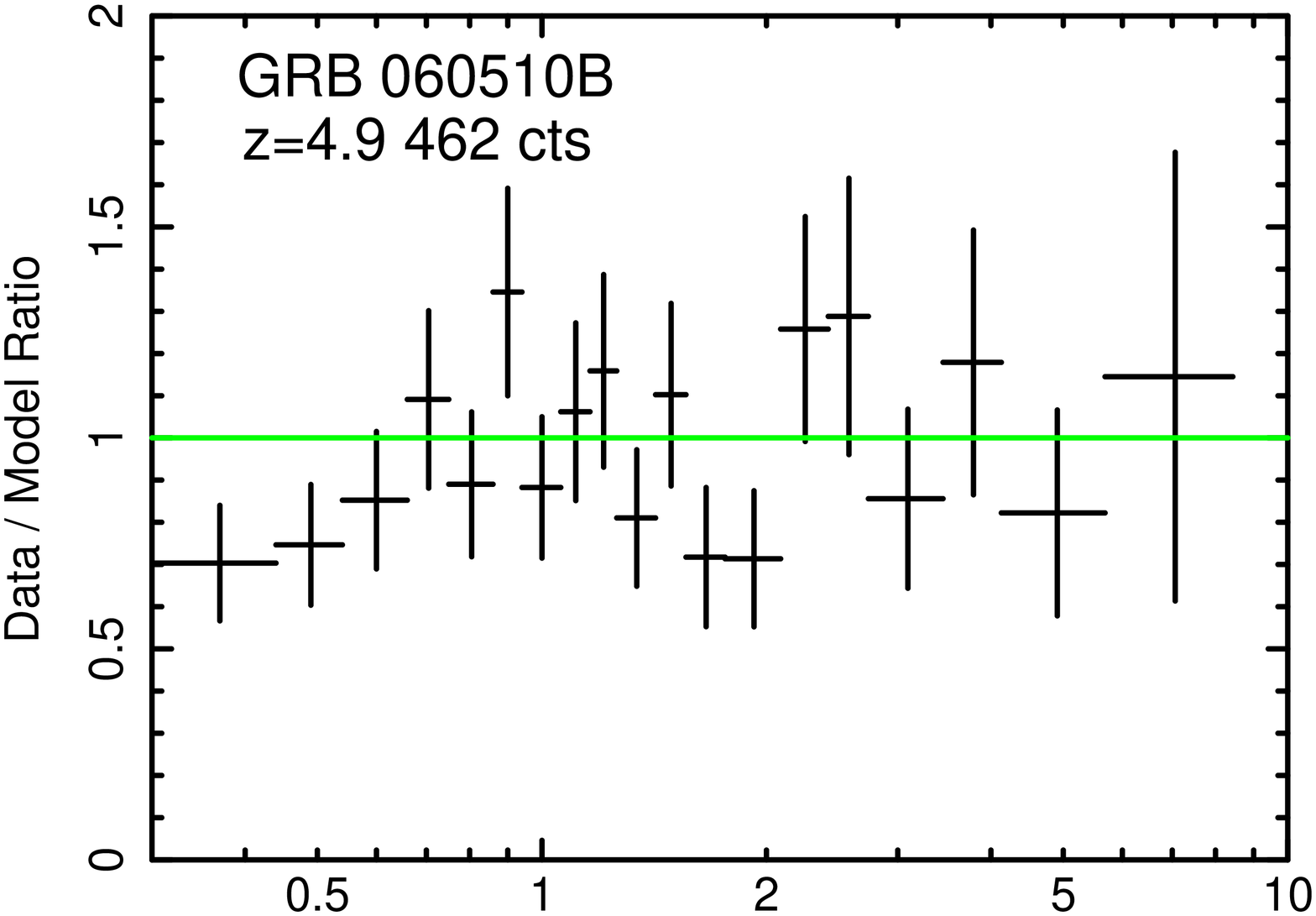}} 
{\includegraphics[angle=-90,width=5cm]{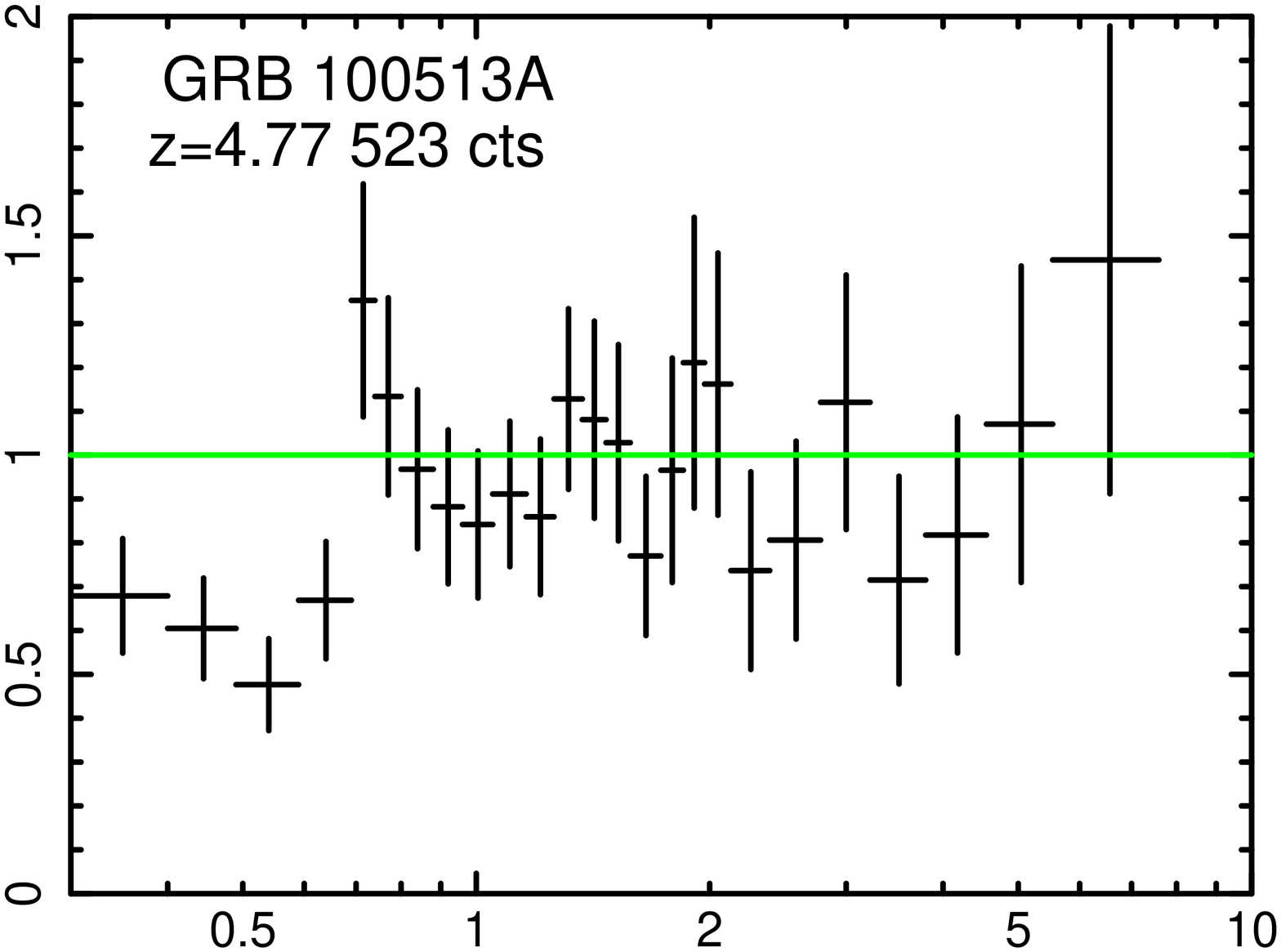}}
{\includegraphics[angle=-90,width=5cm]{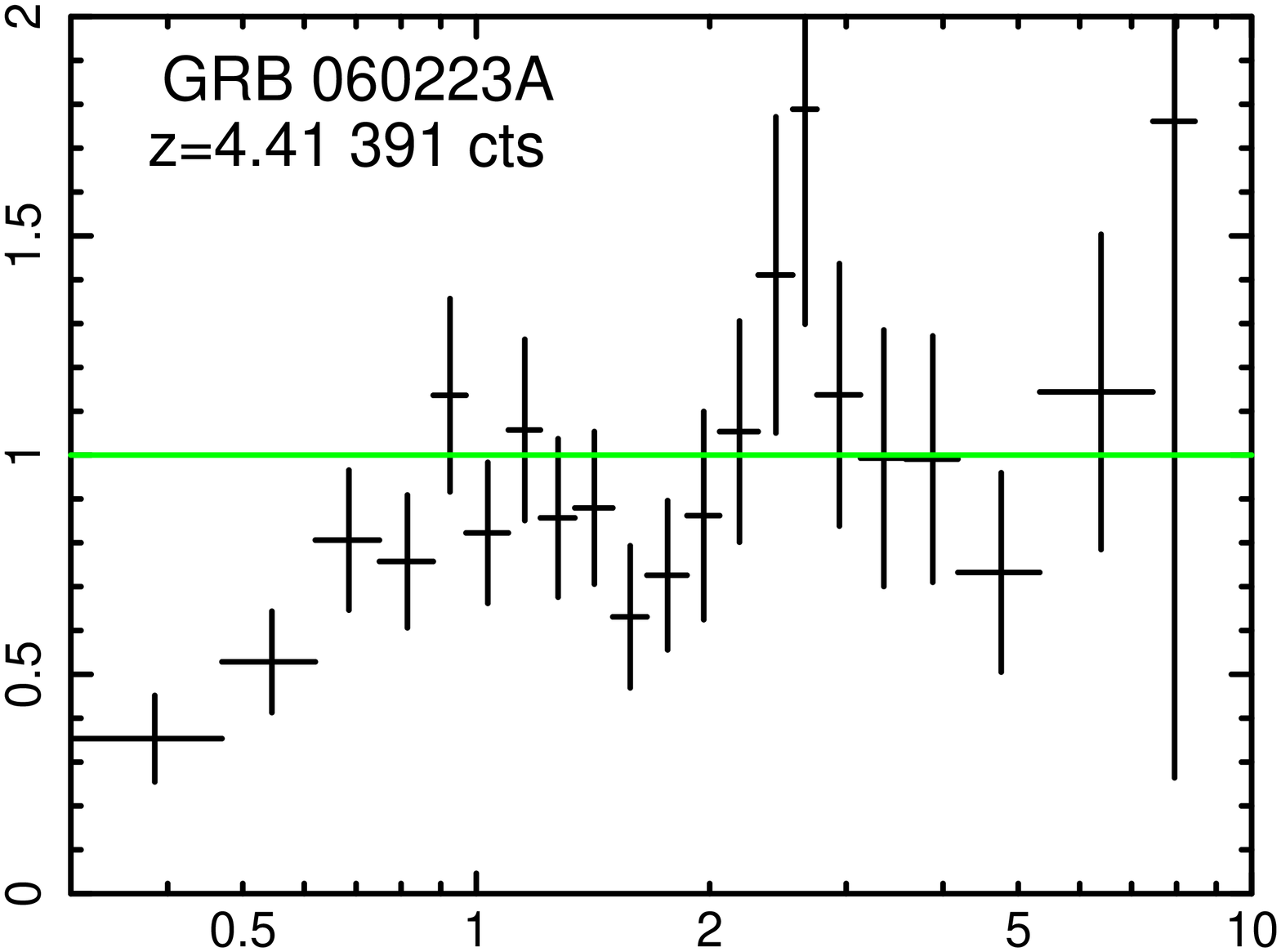}}
{\includegraphics[angle=-90,width=5.3cm]{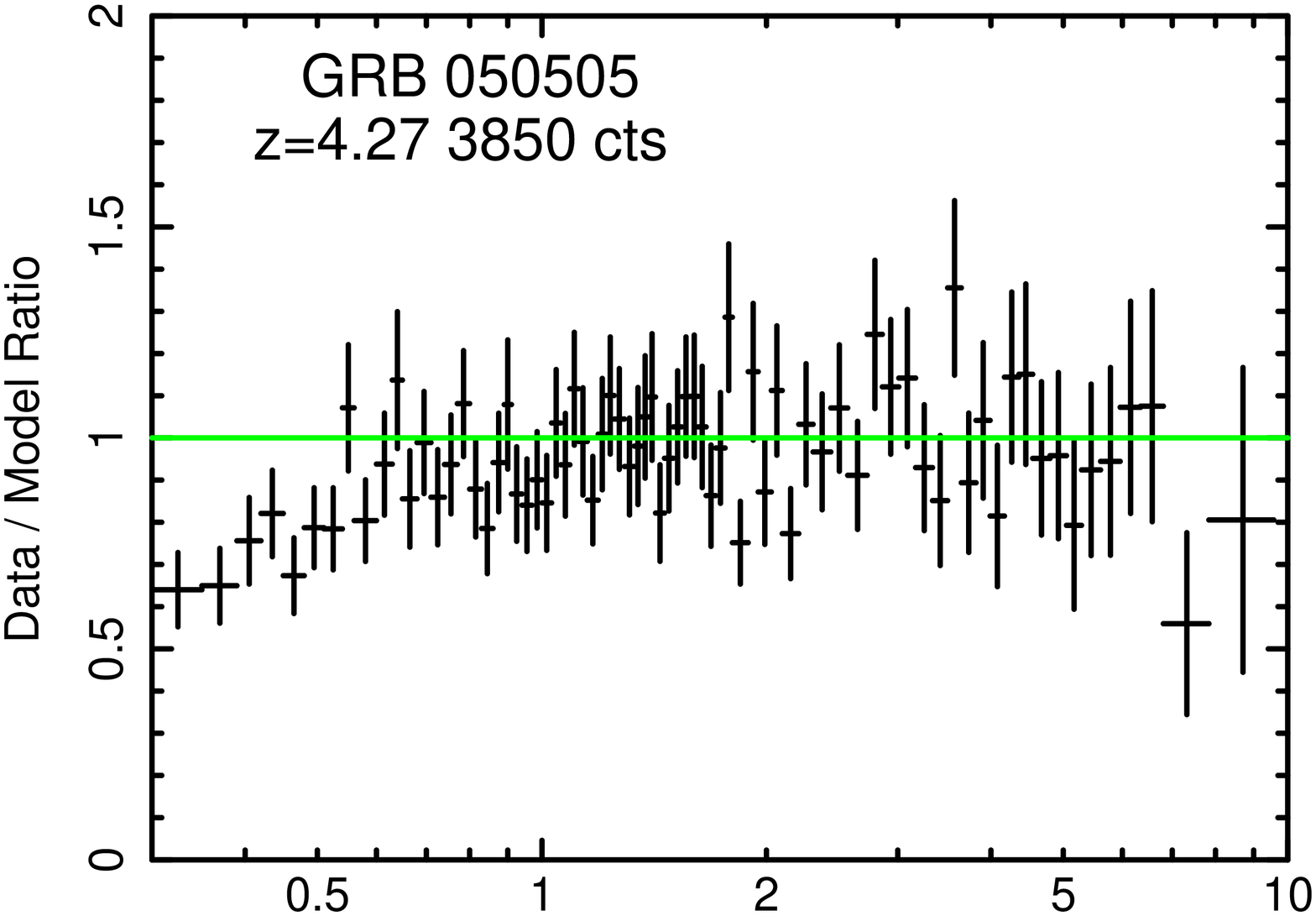}}
{\includegraphics[angle=-90,width=5.cm]{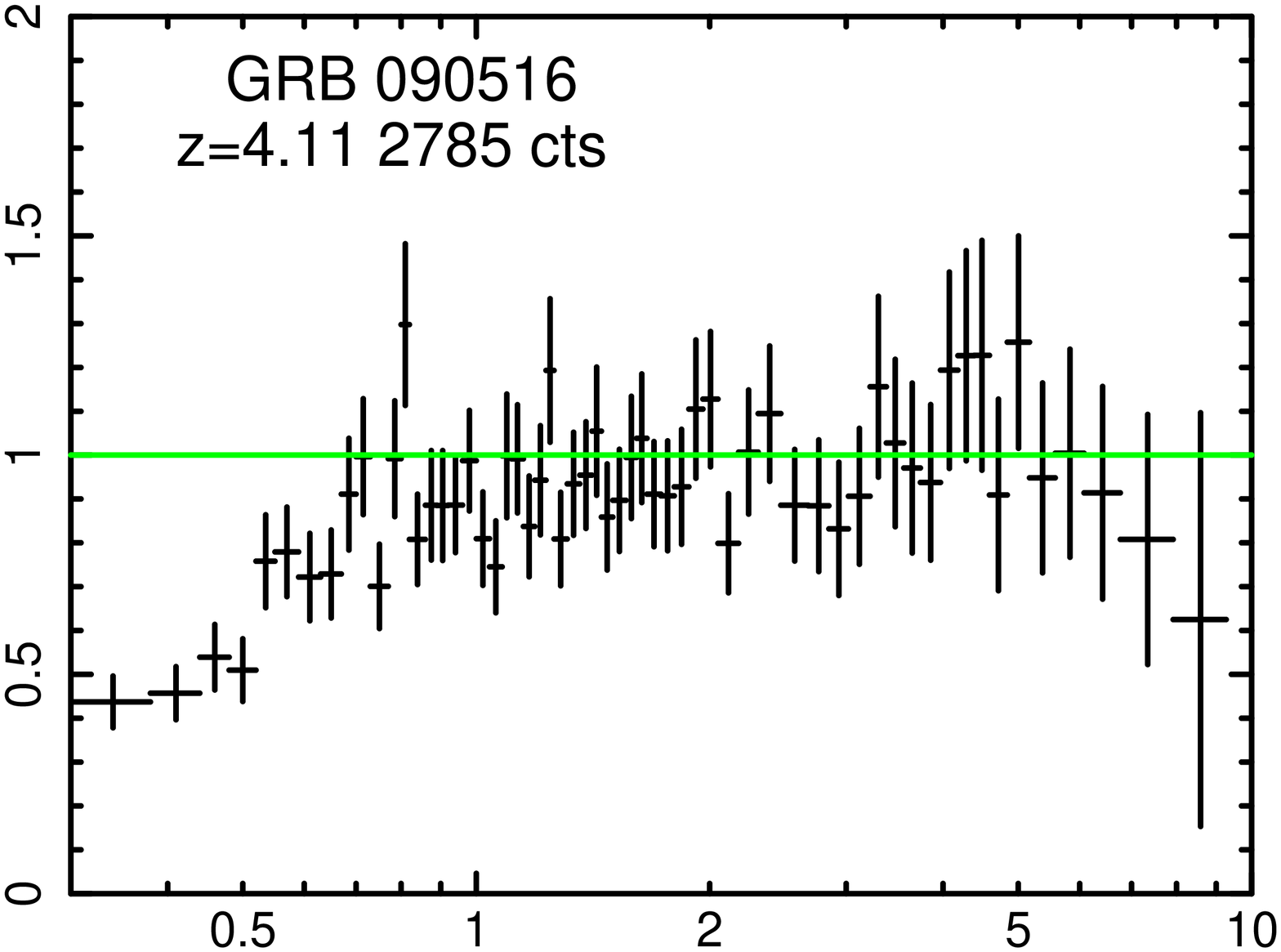}}
{\includegraphics[angle=-90,width=5cm]{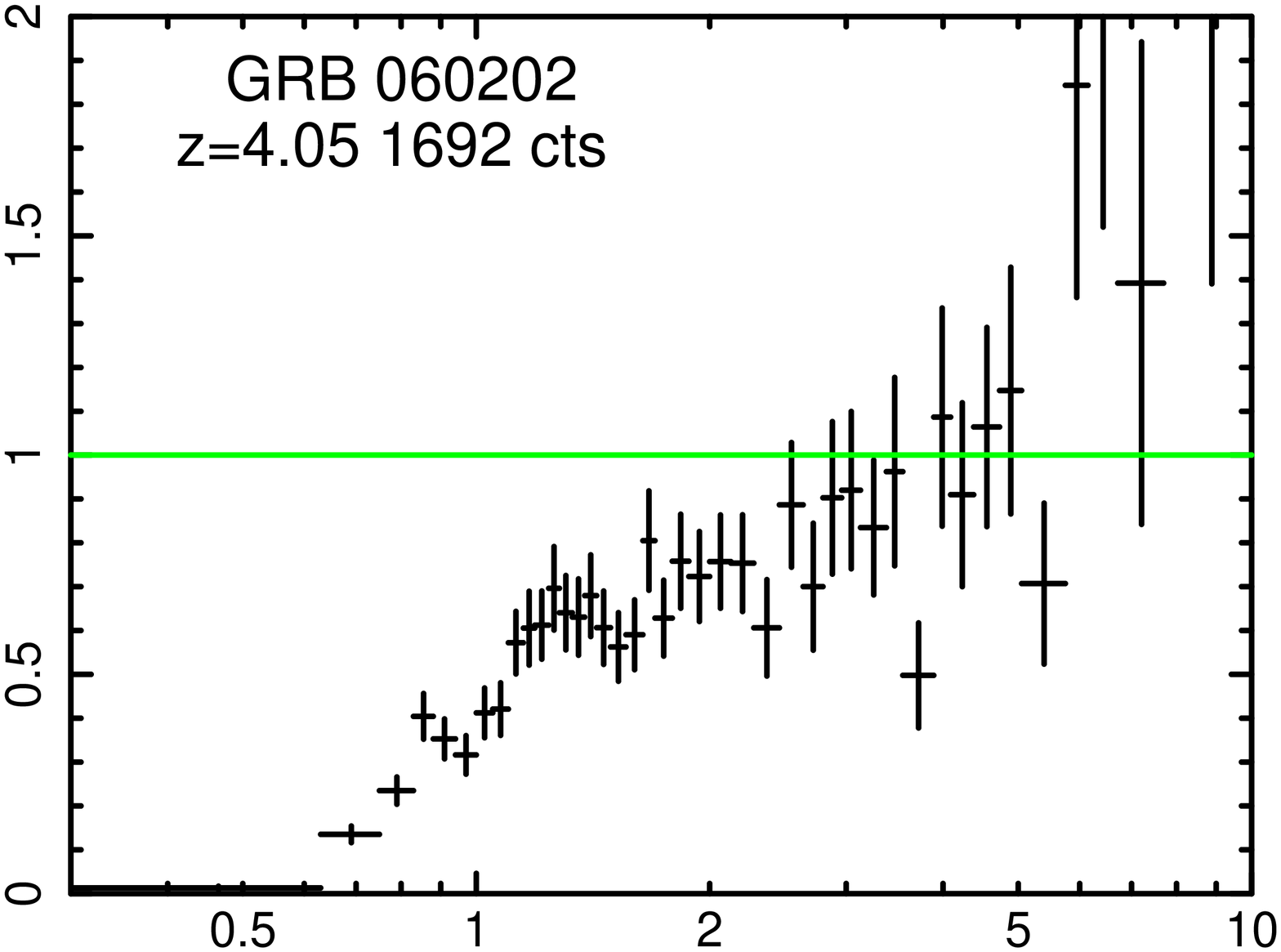}}
{\includegraphics[angle=-90,width=5.3cm]{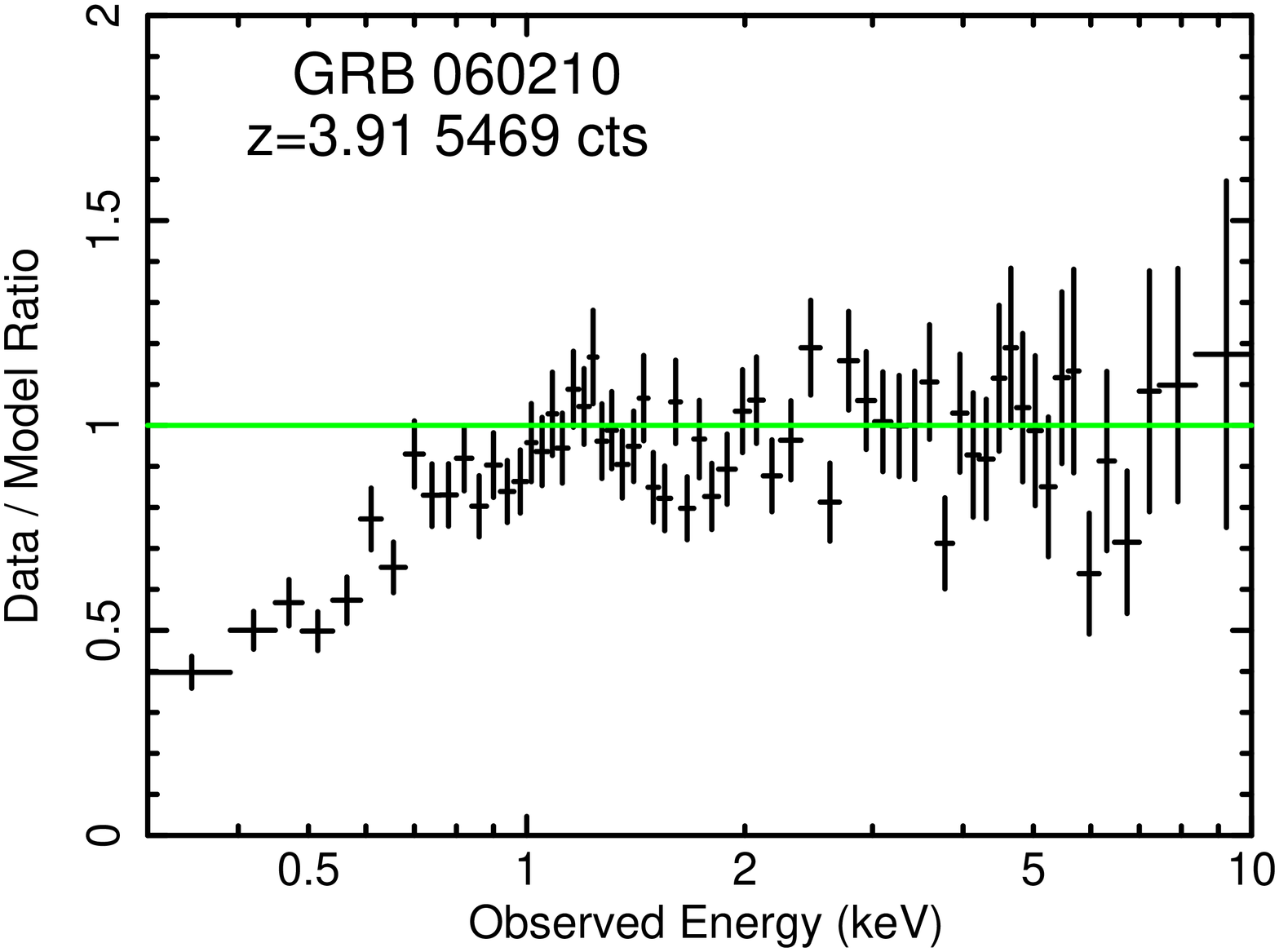}}
{\includegraphics[angle=-90,width=5cm]{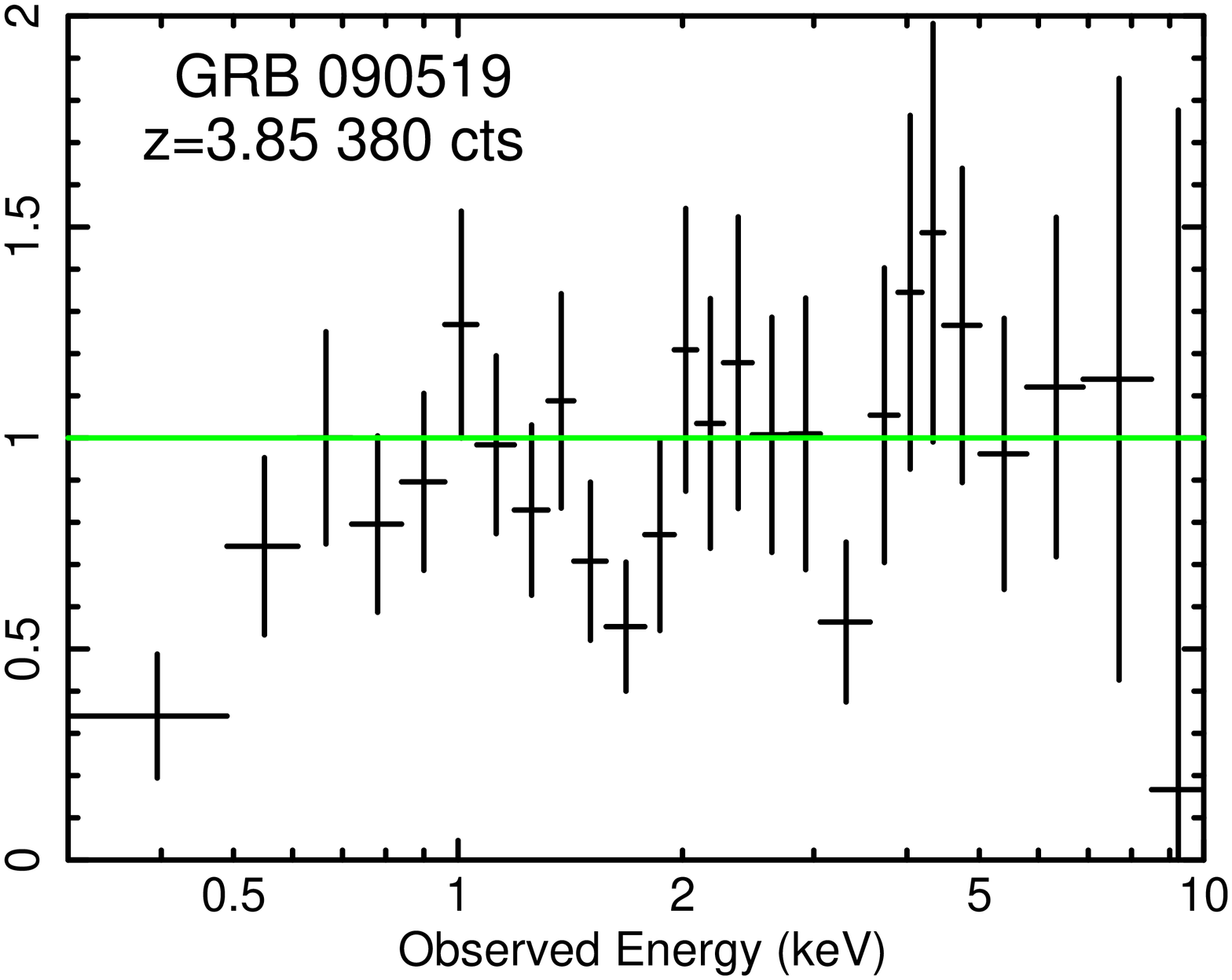}}
{\includegraphics[angle=-90,width=5cm]{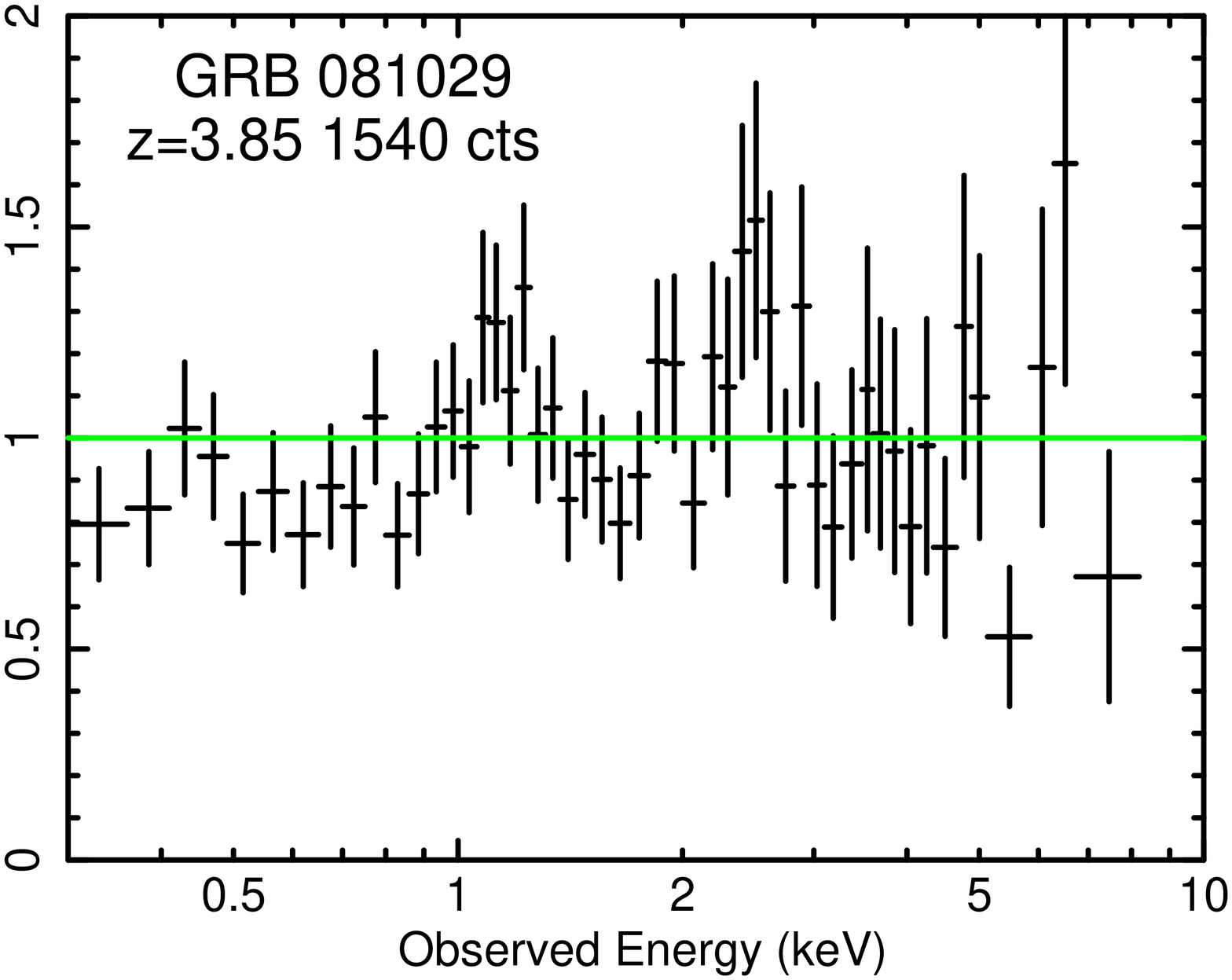}}
\vglue0.0cm
\caption{Data to model ratio plots for the twelve highest-$z$ GRBs with confirmed absorption. 
Data are binned to conveniently represent the extra-galactic transmission functions.
Note the overall similar absorption amplitude irrespective of $z$.}
\label{fig:GRB} 
\end{center}
\end{figure}

The unambiguous opacity at high-$z$ and its apparent independence of $z$ raises the suspicion that the absorption is not intrinsic to the GRB, but is due to a rather uniform foreground. 
We focus on the high-$z$ population where host absorption is strongly quenched (as $\sigma$ sharply declines with $E$, see Sec.~\ref{IGM}).
Therefore, at such high $z$ the possibility that the GRB environment or host galaxy contribute to the observed absorption is small, unless the intrinsic column is of the order of $N_H \sim 10^{22-23}$~cm$^{-2}$, well above the typical galactic column in the local universe.
In order to illustrate the similarities of the absorption spectra irrespective of $z$, in Fig.~\ref{fig:GRB} we plot spectral ratio plots for the twelve highest-$z$ (3.85 -- 8.26) GRBs with absorption confirmed at the 90\% level. 
This sub-sample is drawn from the full \swift /XRT sample shown in Fig.~\ref{fig:tau}, but is not necessarily representative of it.  
For Fig.~\ref{fig:GRB}, each spectrum is fitted to an absorbed power law \citep[with][abundances]{Wilms2000}, after which the excess (extra-galactic) absorption is removed.
The ratio of the model (with only Galactic absorption) to the data, i.e., transmission, is plotted.
The apparent drop of the ratio at low energies ($E < 1$ keV) thus reflects the additional photoelectric non-Galactic absorption towards each GRB. 
GRB\,060202 is exceptional in Fig.~\ref{fig:GRB} with its excessive absorption, and can be seen also in Fig.~\ref{fig:tau} at $1+z = 5.05$ far above the mean GRB opacity.
On the other hand, GRB\,081029 at $1+z = 4.85$ is less absorbed ($\tau \approx 0.16^{+0.14}_{-0.08}$), although, formally, not as well constrained. 
For the most part, absorption sets in at $E<1$~keV, and the transmission at $E=0.5$~keV reaches $\simeq 0.6 - 0.7$, i.e. $\tau \simeq 0.3 - 0.6$. 
The column densities required to produce an opacity $\tau$ that is fixed with $z$ increase roughly as $(1+z)^{2.5}$ due to the corresponding decrease of the photo-ionization cross section with energy, as can be seen in Fig.~\ref{fig:cs} and as explained in more detail below.
These columns then exceed 10$^{23}$cm$^{-2}$ as reported by \citet{Campana2010}.
In the next section, we explore an alternative explanation for the observed constancy of optical depth with $z$.

\section{Soft X-ray opacity of the diffuse IGM}
\label{IGM}

A natural origin of universal, isotropic, X-ray opacity that saturates at high-$z$ is the 
diffuse intergalactic medium (IGM).
In this section, we describe a simple diffuse IGM model, based on well established cosmological parameters, that can explain a tendency to a constant \x\ opacity for high-$z$ sources. 
Since the photo-ionization cross section per H atom in the photon energy range of 0.5~keV $< E < $ 10~keV scales roughly as $\sigma (E) \propto E^{-2.5}$ (Fig.~\ref{fig:cs}), for a redshifted absorber at a fixed observed energy $E , \sigma (E,z) \propto (1+z)^{-2.5}$.
Also, since the absorption at \x\ energies is dominated by metals, one can assume $\sigma (E,z,Z_\odot) \propto Z_\odot \sigma (E,z) $ scales approximately with the IGM metallicity $Z_\odot$ (in solar units) that can evolve with redshift as $Z_\odot (z) = Z_0 \eta(z)$. We can thus write for the IGM photo-electric optical depth

\begin{eqnarray} 
\label{tauIGM} 
\tau_{IGM}(E,z,Z_\odot) = \int_0^z n_H(z')\sigma (E,z',Z_\odot) c \left( \frac{dt'}{dz'} \right) dz' \approx \nonumber \\
\frac{n_0 c Z_0}{H_0} \sigma (E,0) \int_0^z  \frac{(1+z')^3 \eta (z') dz'}{(1+z')^{2.5}(1+z')\sqrt{(1+z')^3\,\Omega_M +\Omega_\Lambda}}
\end{eqnarray}

\noindent where $n_0$ and $Z_0$ are the $z = 0$ mean IGM hydrogen number density and metallicity (in solar units), $c$ is the speed of light, and $H_0$ is the Hubble constant.
$\Omega_M$ and $\Omega_\Lambda$ are, respectively, the present-day matter and dark energy fractions of the critical energy density of the Universe.
Even with no metallicity evolution, i.e., $\eta(z) \equiv 1$, the differential optical depth $d\tau / dz' \to (1+z')^{-2}$ at $z\gg 1$, and the integral in Eq.~\ref{tauIGM} thus saturates at the high-$z$ limit.
This is starkly different from the more intuitive continuous increase of optical depth with $z$, e.g., for Compton scattering and line absorption.
The likely decrease of metallicity with $z$ makes $\tau_{IGM}(E,z)$ saturate even faster with $z$.
The behavior of $\tau_{IGM}(E,z)$ without metallicity evolution and with $\eta (z) \propto (1+z)^{-2}$ is shown in Fig.~\ref{fig:igm}, overlaid on the measured opacities.
These curves are {\it not} fits to the data, but are plotted to give a rough idea of how the IGM contribution to the opacity saturates at high $z$.

The mean density of hydrogen in the IGM is $n_0 \approx 0.67\,\Omega_b\, (3\, H_0^2/8\, \pi\, G\, m_H)$ = 1.7$\times 10^{-7}$ cm$^{-3}$, where $\Omega_b = 0.045$ is the Universe's baryon energy fraction \citep{Komatsu2010}, $G$ is the gravitational constant, and $m_p$ the hydrogen mass.
The pre-factor of 0.67 comes from the fact that $\sim 74\%$ of the Universe mass is in hydrogen atoms $\sim$90\% of which reside in the IGM \citep{Fukugita2004}.
The Hubble constant is $H_0 = 71\,{\rm km\, s^{-1}\, Mpc^{-1}}$ and the cross section at 0.5~keV $\sigma$(0.5~keV,0) = $6.22\times 10^{22}$ cm$^{-2}$.
For $\eta(z) \equiv 1$ (no metallicity evolution), and using the standard cosmological parameters $\Omega_M = 0.27$ and 
$\Omega_\Lambda\ = 0.73$ \citep{Komatsu2010}, the integral in Eq.~\ref{tauIGM} up to $z = 10$ attains a value of $\sim$1.4.
It is interesting that this estimate results in the simple asymptotic expression $\tau_{IGM}(0.5\, \mathrm{ keV},z \gg 1) \approx 2Z_0$.
In other words, a reasonable $z = 0$ metallicity of $Z_0 \sim 0.2$ could explain the measured asymptotic value of $\tau \sim 0.4$. 

There are two strong, oversimplifying assumptions in this estimate that have to do with the metallicity and the ionization of the IGM.
First, there is most definitely a metallicity evolution in the IGM \citep[e.g.,][]{simcoe04}.
If one assumes a universal metallicity evolution of $Z_\odot (z) = Z_0(1+z)^{-k}$, the saturation of $\tau_{IGM}(E,z)$ occurs even at lower $z$ than without evolution.
 In that case, for $z' \gg 1$, $d\tau / dz' \to (1+z')^{-2-k}$, and the integral in Eq.~~\ref{tauIGM} is approximately (1+k) times smaller. In other words, the corresponding expression is
 
 \begin{equation}
 \tau_{IGM}(0.5\, \mathrm{ keV},z \gg 1) \approx\ 2Z_0/(1+k)
 \label{tauk}
 \end{equation}
 
\noindent For example, if one adopts  for the IGM the metallicity trend of $Z_\odot \approx (0.54 \pm\ 0.10)(1 + z)^{-1.25\, \pm\ 0.25}$ (i.e., $Z_0 \approx 0.5$ and $k \approx 1.25$) observed for Fe in the \x\ emitting gas of galaxy clusters up to $z \approx 1$ \citep[][]{Balestra2007}, the above relation suggests $\tau_{IGM}(0.5\, \mathrm{ keV},z \gg 1) \approx 0.45 \pm 0.15$, which is consistent with both the GRB mean value (above) and with the quasar results (next section). 
Alternatively, $Z_\odot \approx (0.2 \pm\ 0.1)(1 + z)^{-1}$ derived for Fe from the measured supernova rate (Graur et al. 2011, private communications) suggests $\tau_{IGM}(0.5\, \mathrm{ keV},z \gg 1) \approx 0.2 \pm 0.1$, slightly less than the clusters or the above estimated GRB values.

The second caveat is the ionization correction, which is more difficult to deal with, as the ionization state of the diffuse IGM is not directly observed and is a matter of ongoing debate \citep[e.g.,][]{bolton07}.  
The above estimates are all based on the photo-ionization cross section of a neutral absorber.  
An ionized absorber would have a somewhat lower cross section at \x\ energies, and therefore a larger column density would be required to produce the same $\tau$. Calculating cross sections for different ionization states and different metallicities in the IGM is beyond the scope of this paper.
However, in Fig.~\ref{fig:cs} we plot  separately the contributions to the photo-ionization cross section of H and He, and that of the metals with solar composition. 
It can be seen that the metals are dominated by H and He below the O edge at 0.54~keV where they overtake the H and He contribution.
For redshifted absorption it is thus the metal contribution that mostly determines $\tau$(0.5 keV) unless the metallicity is radically sub-solar.
The pure metal contribution can also be thought of as an upper limit to the cross section for ionized plasma in which H and He are totally ionized (although in reality the metal contribution also somewhat decreases with ionization).
A $z = 1$ absorber in which H and He are totally transparent (ionized) would still retain more than 70\% of its opacity at 0.5~keV (observed) mostly due to C and O, 
and that fraction of course needs to be further scaled with the metallicity.

To summarize this section, the mean high-$z$ opacity for the diffuse IGM 
within the standard cosmological model can explain the soft \x\ opacity measured from spectra of high $z$ GRBs (and quasars) as long as the gas is not too highly ionized, and as long as the metallicity is a reasonable fraction of the solar value and does not decrease too quickly with redshift.
The low ionization is possible since the truly diffuse IGM suffered less gravitational collapse than the denser line absorbing IGM systems that have been heated to $\sim 10^6$\,K
\citep{dave01}.
The metallicity of 0.2 -- 0.4 solar required to explain the absorption is actually lower than the Fe abundance observed in the hot gas of galaxy clusters up to $z = 1.3$ \citep{Balestra2007, maughan08}, but slightly higher than that implied by supernova rates (Graur et al. 2011) and
the abundances observed in IGM filaments towards nearby $z < 0.4$ metal absorbers \citep[$Z_\odot \sim 0.1$,][]{danforth08}.
Recall that most of the IGM column is accumulated up to $z = 1 - 2$ (Eq.~\ref{tau}, Fig.~\ref{fig:igm}), so even a sharp drop in metallicity beyond those redshifts does not change our conclusions.
In fact, the mean metallicity in damped Lyman-$\alpha$ (DLA) absorbers up to 
$z < 4$ \citep{Savaglio2009, Kaplan2010} is also consistent with these values.
although the scatter in metallicity in DLAs \citep{Prochaska2003, Savaglio2009, Kaplan2010}, 
and in galaxy clusters \citep{Balestra2007}, at any given $z$ is quite large.


 
\subsection{Host absorption}

It is impossible to rule out a cosmological evolution of the host (GRB environment or galaxy) column density that follows the values found by \citet{Campana2010} and reaches $N_H \sim 10^{23}$ cm$^{-2}$.
On the other hand, at high redshift it is reasonable to assume that the contribution of the host galaxy to the opacity at 0.5 keV is small because of the redshift effect and the photo-ionization cross-section energy dependence $\sigma (E) \sim E^{-2.5}$ discussed above.
At low redshift the opacity of the host certainly is dominant, but suffers from a large spread as is seen in Fig.~\ref{fig:tau}, see also \citet{Campana2010}.
This spread impedes any concrete conclusions regarding the hosts based only on their \x\ absorption properties.
Such a spread is expected from the variety of  host galaxies, GRB environments, GRB locations within the galaxies, and lines of sights. 
In Fig.~\ref{fig:igm} we include a (high) mean host absorber of $N_H = 3\times 10^{21}$cm$^{-2}$, whose contribution to $\tau$(0.5 keV) is significant at low redshift, but becomes less important than the IGM at $z \sim 1.5$.
At high $z$, where the host contribution gradually becomes negligible, 
the spread in opacity values indeed becomes smaller,
and the theoretical IGM opacity curve, thus, well describes the observed values. 
Since even for high-$z$ sources, much of the IGM opacity is due to the gas up to $z < 2$ (see Eq.~\ref{tauIGM}), a clumpy IGM \citep[e.g.,][]{wiersma11} can also cause a spread in opacity at high $z$.
 
%

\section{Observed-frame opacity of \x\ selected high-$z$ quasars}
\label{quasars}

\begin{figure}[pt!]
{\includegraphics[width=13cm, angle=-90]{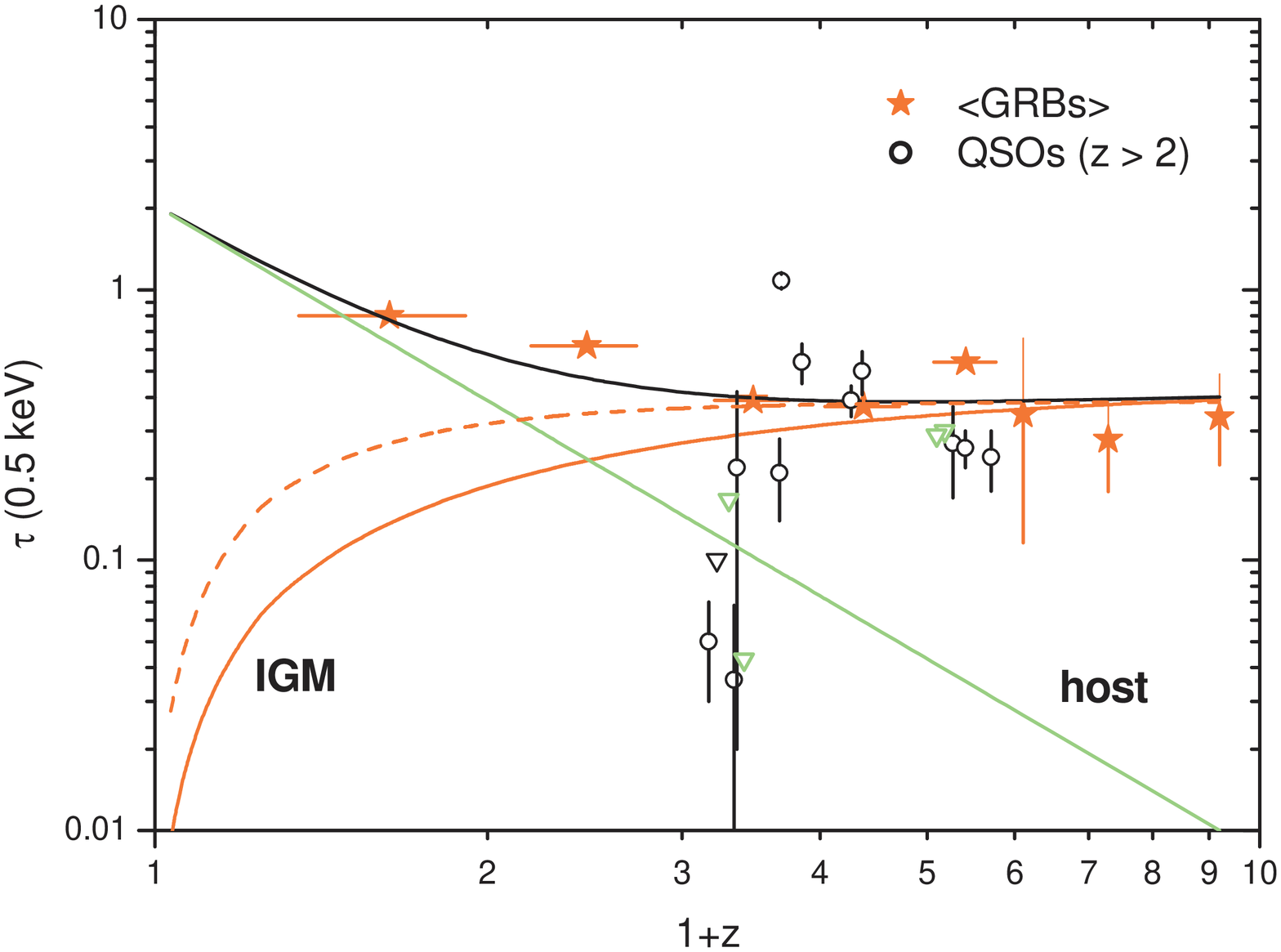}}
\caption{Optical depth $\tau$ at 0.5 keV as a function of redshift. 
Mean \swift /XRT GRB values are denoted by red stars (c.f., Fig.~\ref{fig:tau}). 
Black (green) circles (triangles) represent the RLQs (RQQs) measured with \xmm .
Upper limits are denoted by down pointing triangles (RQQs are all upper limits).
Theoretical IGM and host-galaxy contributions (Sec.~\ref{IGM}) to the total opacity (black line) are shown separately. 
The IGM theoretical contribution is scaled to approach $\tau = 0.4$ at high $z$. 
The dashed line represents the IGM contribution if its metallicity evolved as $Z_\odot (z) = Z_0 (1+z)^{-2}$.  }
\label{fig:igm} 
\end{figure}

We wish to further explore whether soft \x\ absorption of high-$z$ GRBs is due to the peculiar environment of the GRBs, or whether there is a significant intervening IGM contribution.
A natural place to check for IGM absorption is with high-$z$ steady X-ray sources, namely quasars.
For this, we searched the \xmm\ archive for all quasars with $z > 2$ for which the \pn\ camera recorded more than 5000 source photons, to ensure high S/N. We found the 13 quasars listed in Table~\ref{tab:RLQ}, most of which were already individually reported in the literature.
We used the standard \pn\ pipeline products and fitted all the quasar spectra in the exact same manner as the GRB spectra, namely with a host ($z$) absorber.
The column density $N_H(z)$ obtained from this measurement was then translated into the optical depth $\tau$(0.5 keV) according to Eq.~\ref{tau*}.
No abundance correction was needed (Eq.~\ref{tau}) as we used the abundances of \citet{Wilms2000} for both the Galactic and host absorbers.
The best-fit $\tau$(0.5 keV) values are listed in Table~\ref{tab:RLQ} and plotted in Fig.~\ref{fig:igm} and can be seen to compare well with the mean GRB opacities.
Quoted and plotted errors correspond to the 90\% confidence limits.
When more than one \xmm\ observation is available, we fit all observations simultaneously with a single absorbing column. 

\begin{figure}[pt!]
\begin{center}
\vglue0.0cm
{\includegraphics[angle=-90,width=5.3cm]{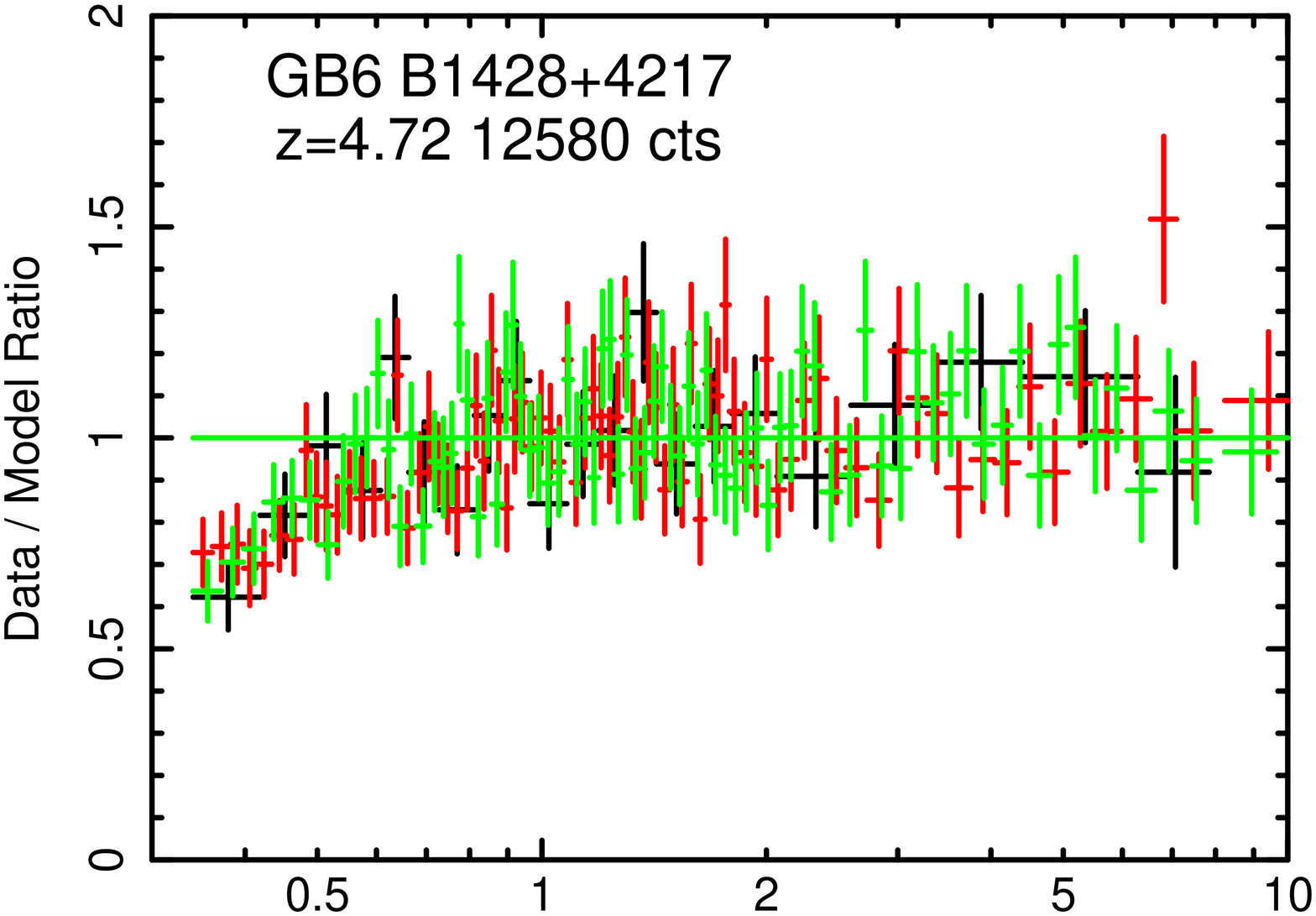}}
{\includegraphics[angle=-90,width=5cm]{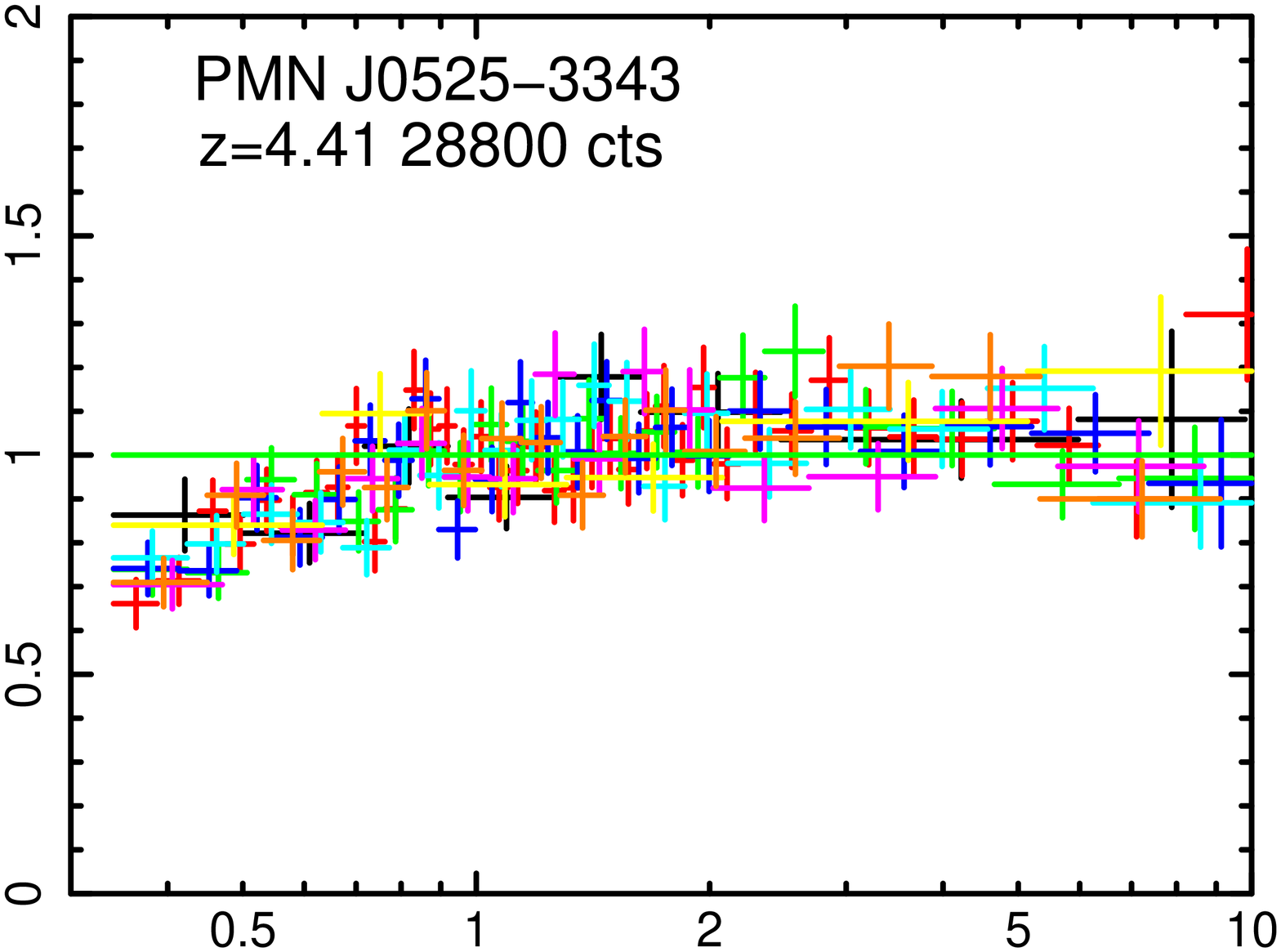}}
{\includegraphics[angle=-90,width=5cm]{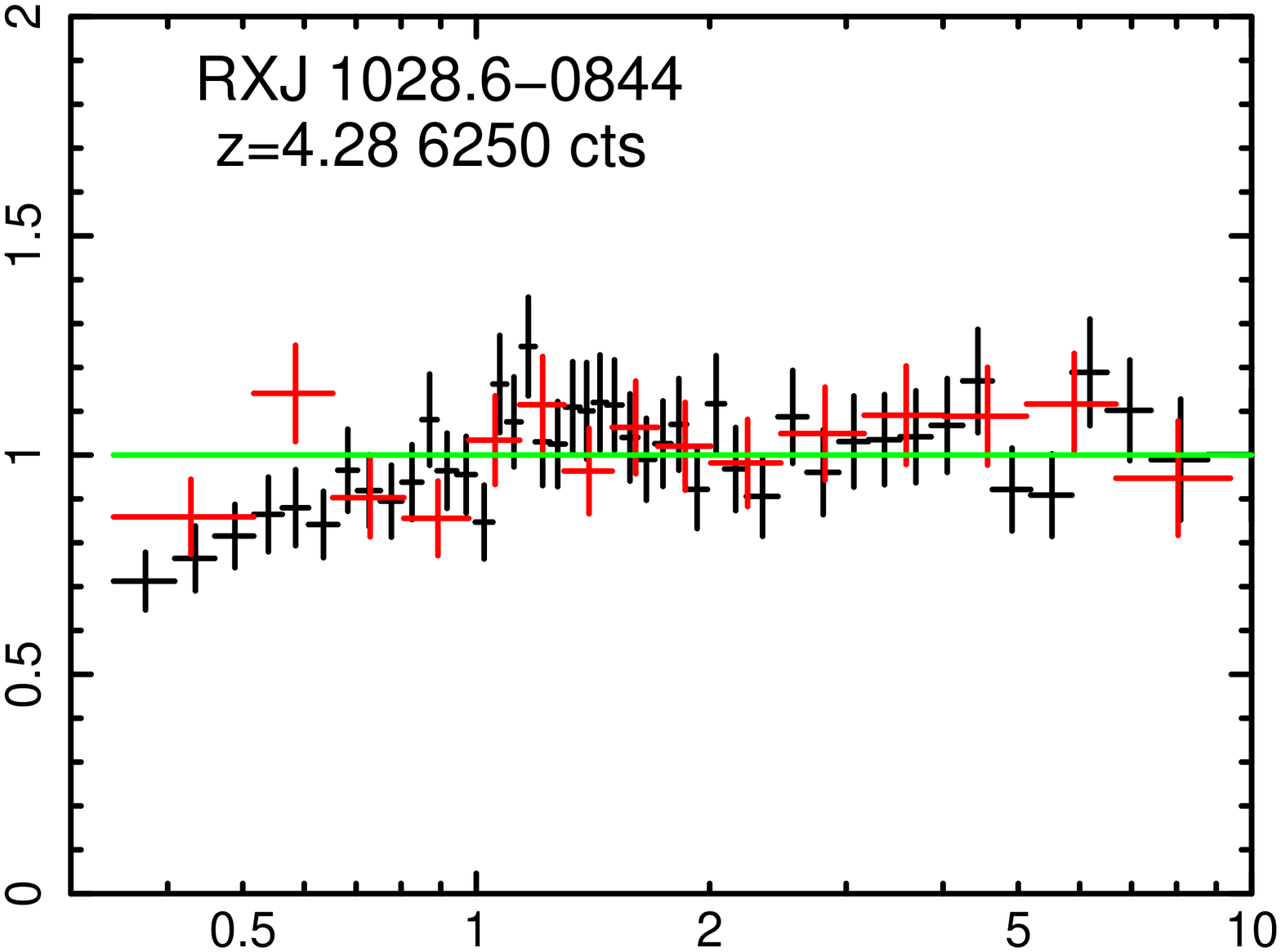}}
{\includegraphics[angle=-90,width=5.3cm]{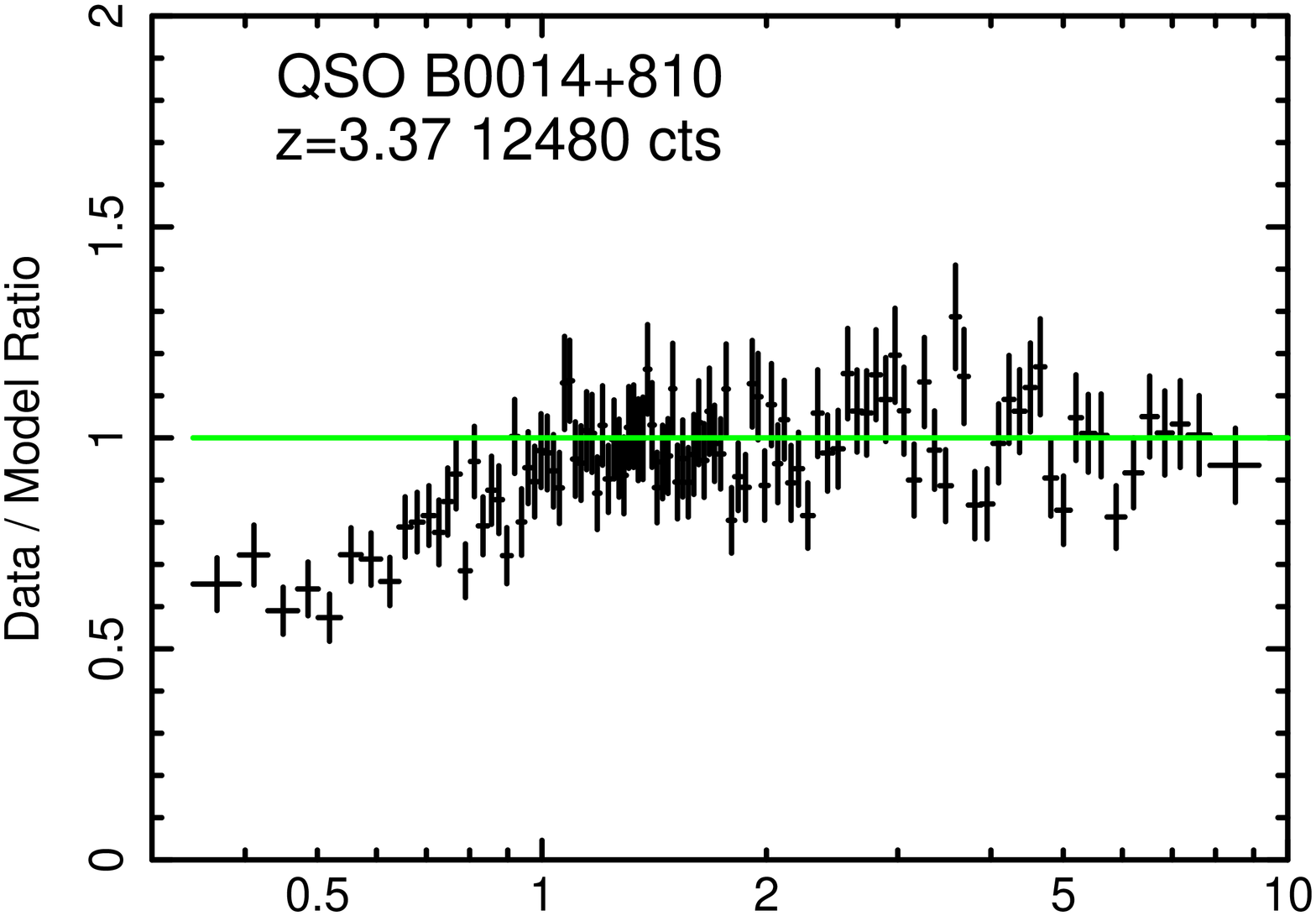}}
{\includegraphics[angle=-90,width=5cm]{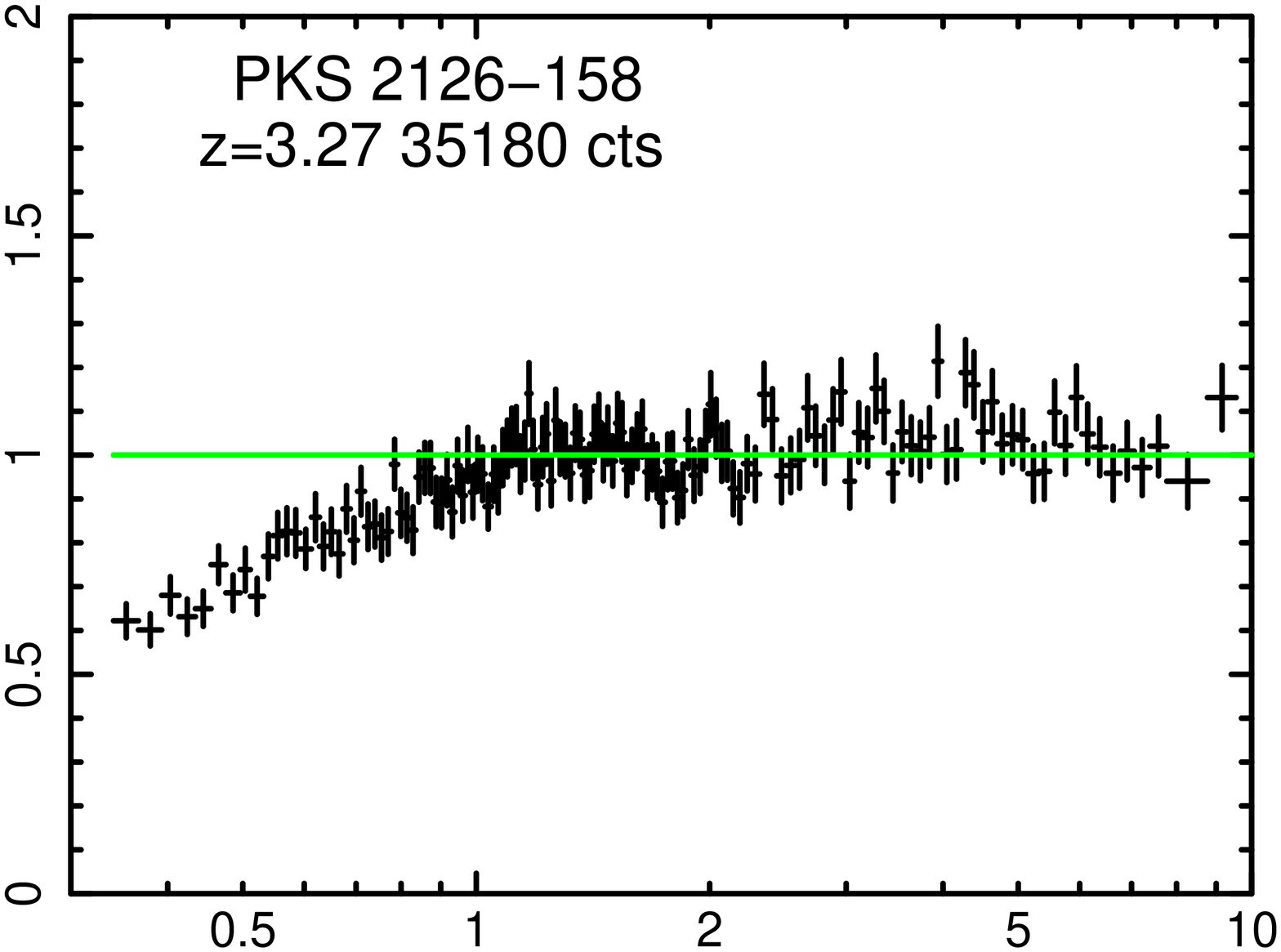}}
{\includegraphics[angle=-90,width=5cm]{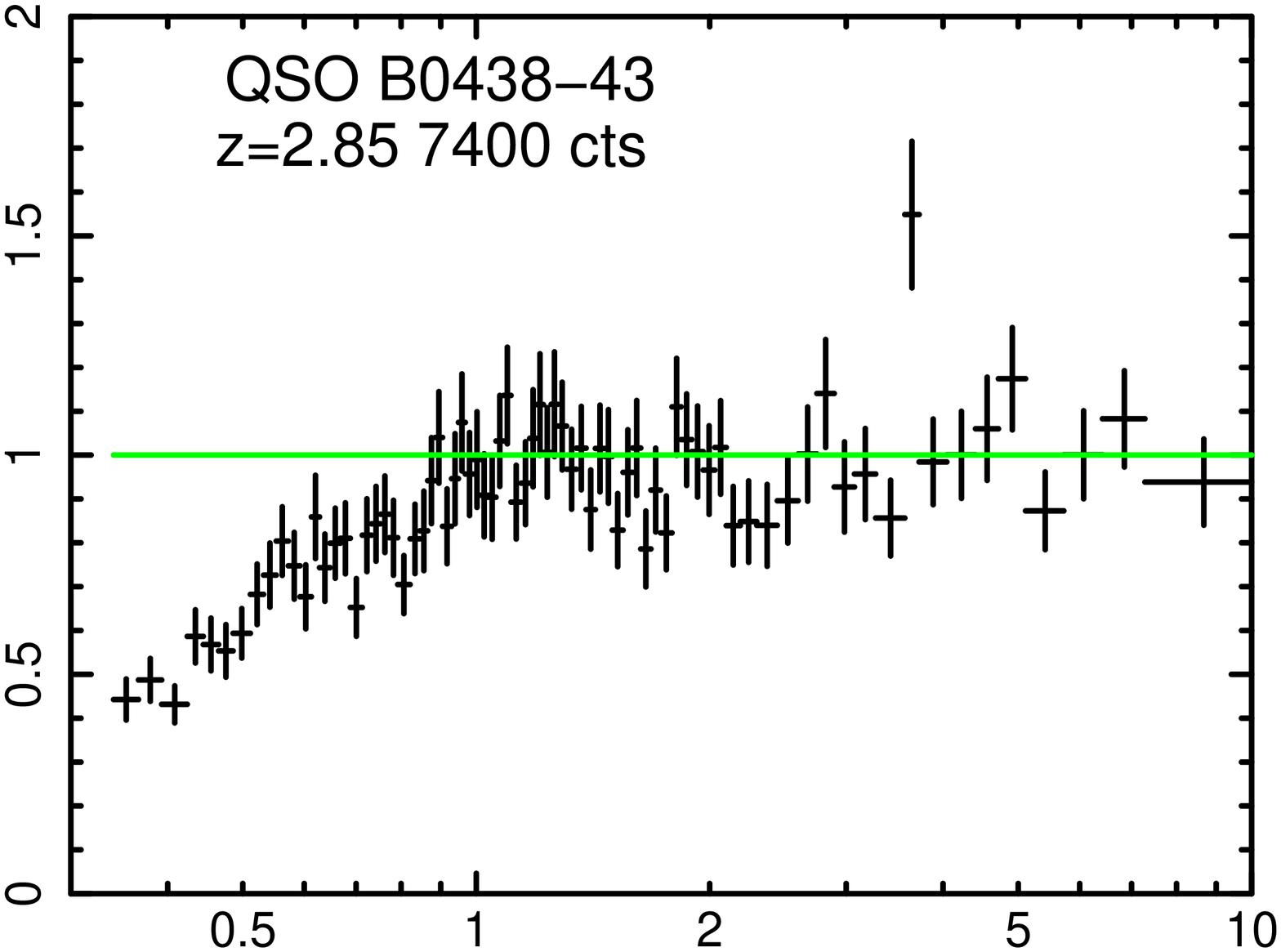}}
{\includegraphics[angle=-90,width=5.3cm]{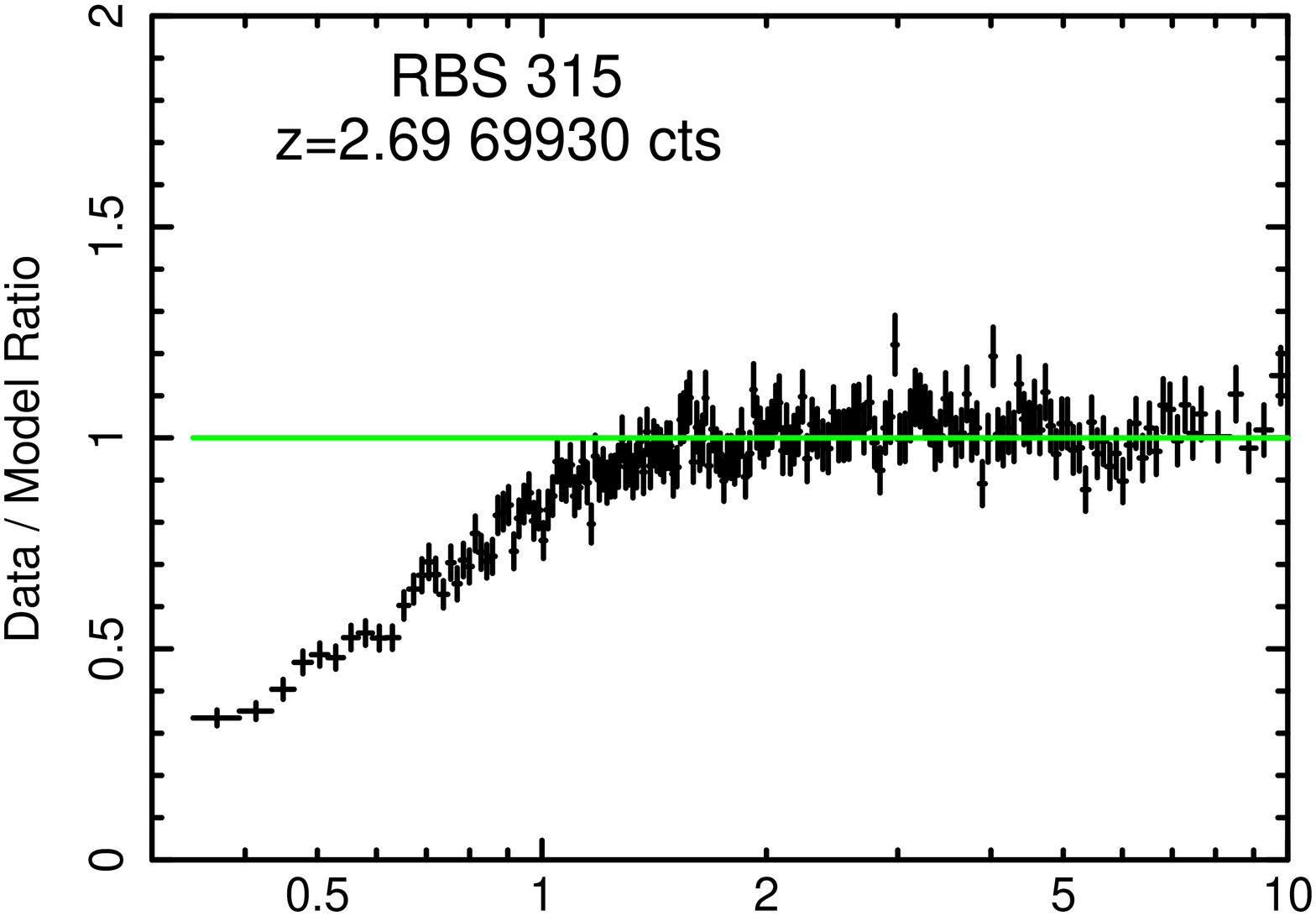}}
{\includegraphics[angle=-90,width=5cm]{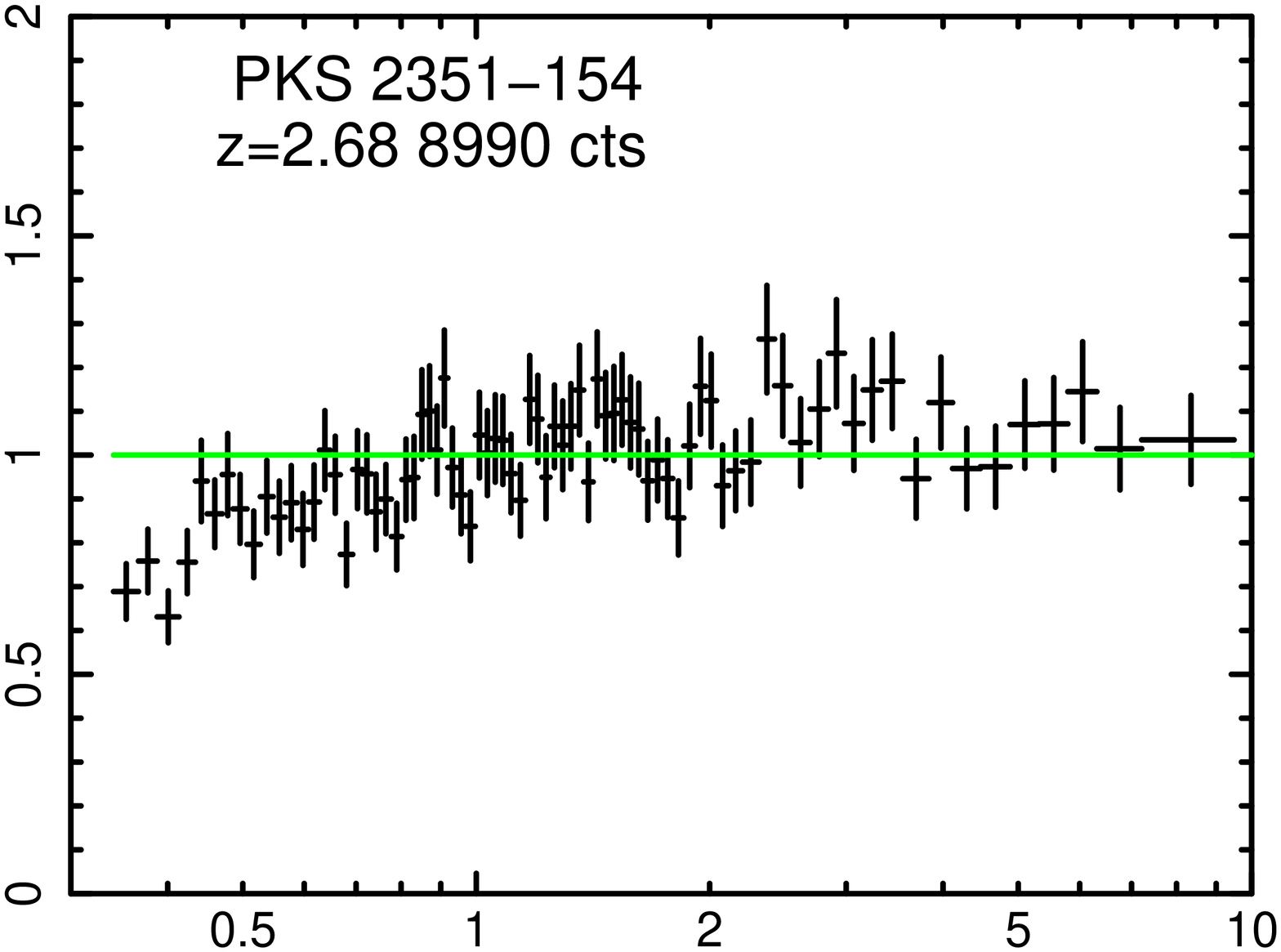}}
{\includegraphics[angle=-90,width=5cm]{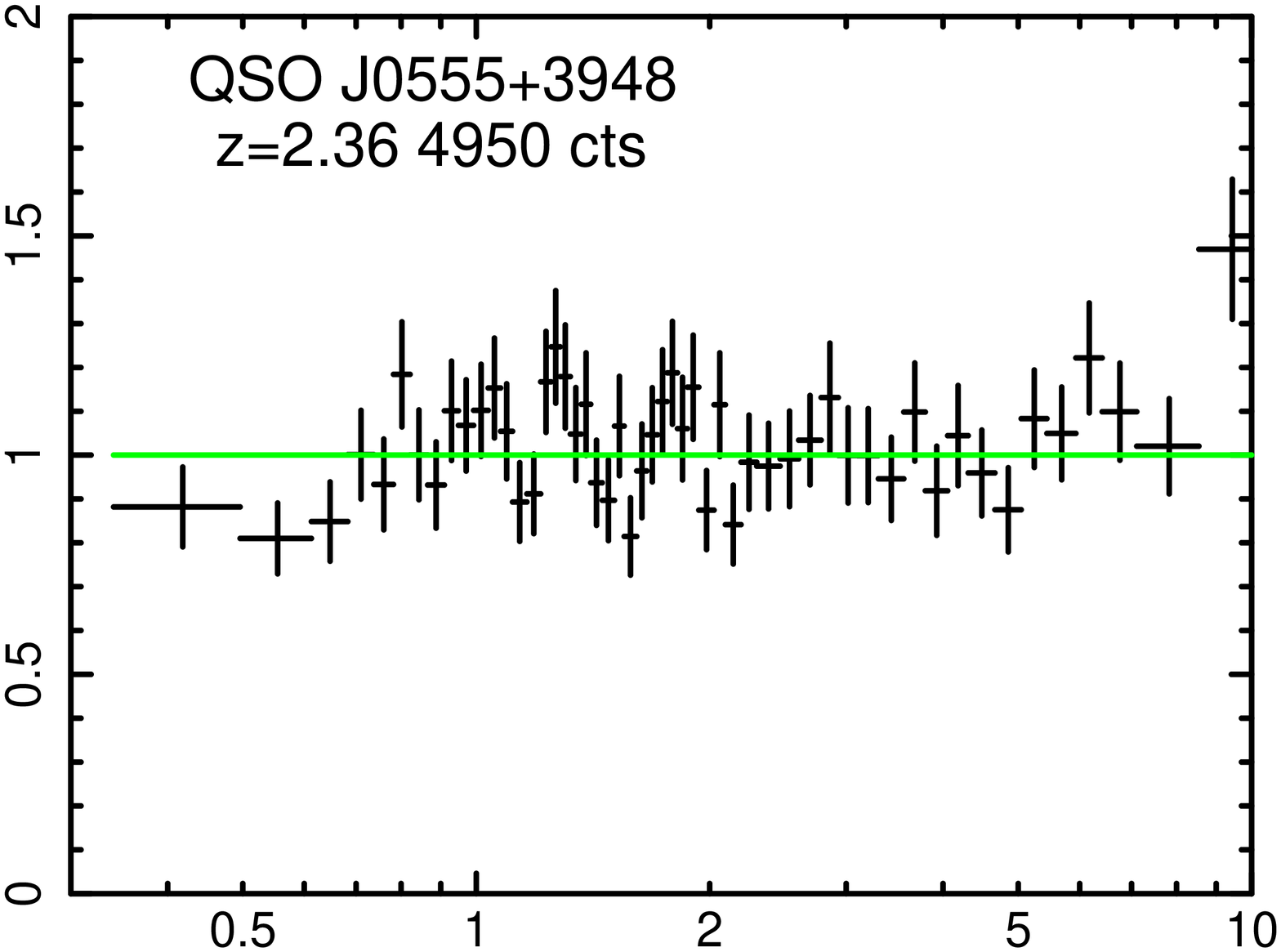}}
{\includegraphics[angle=-90,width=5.3cm]{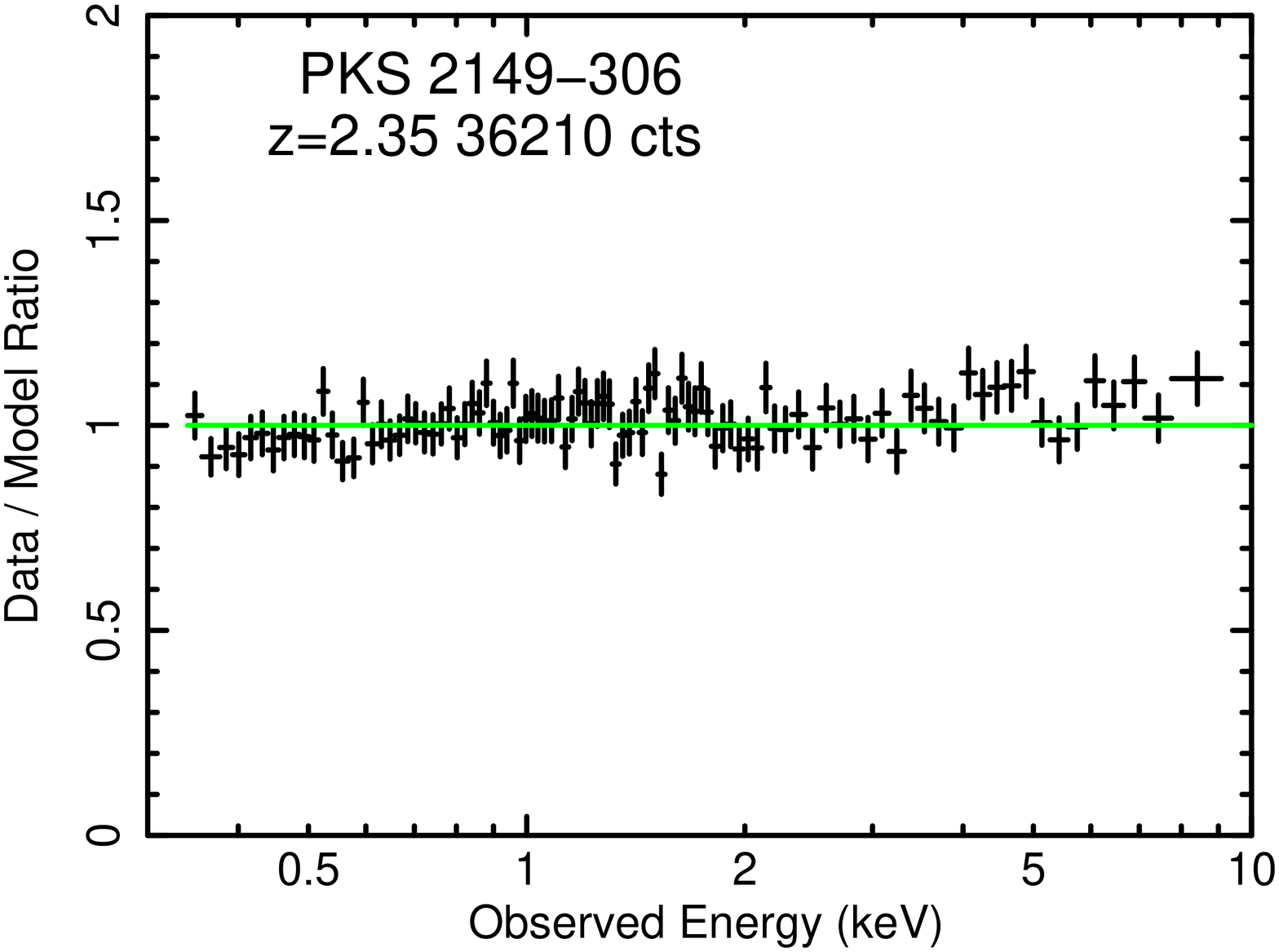}}
{\includegraphics[angle=-90,width=5cm]{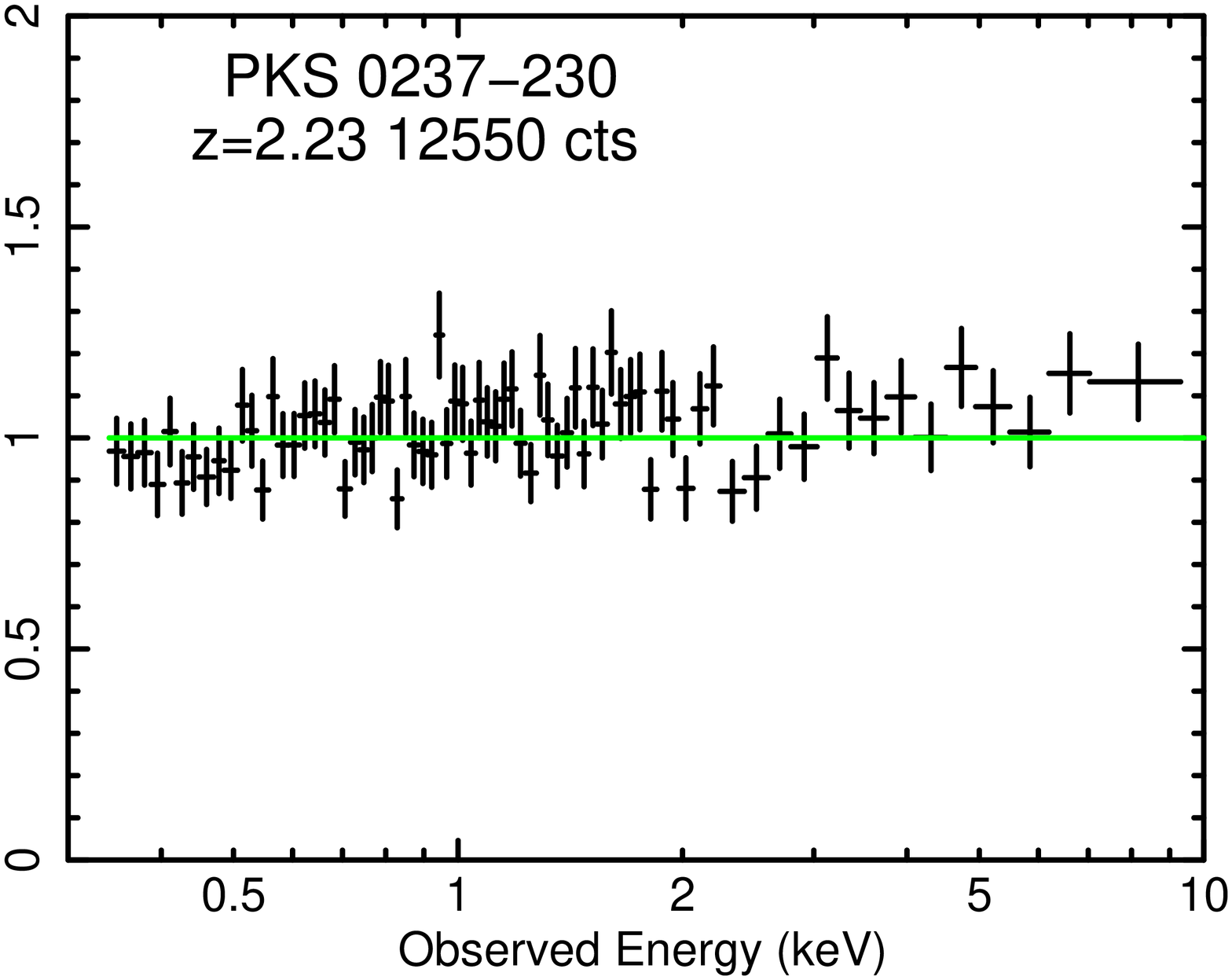}}
{\includegraphics[angle=-90,width=5cm]{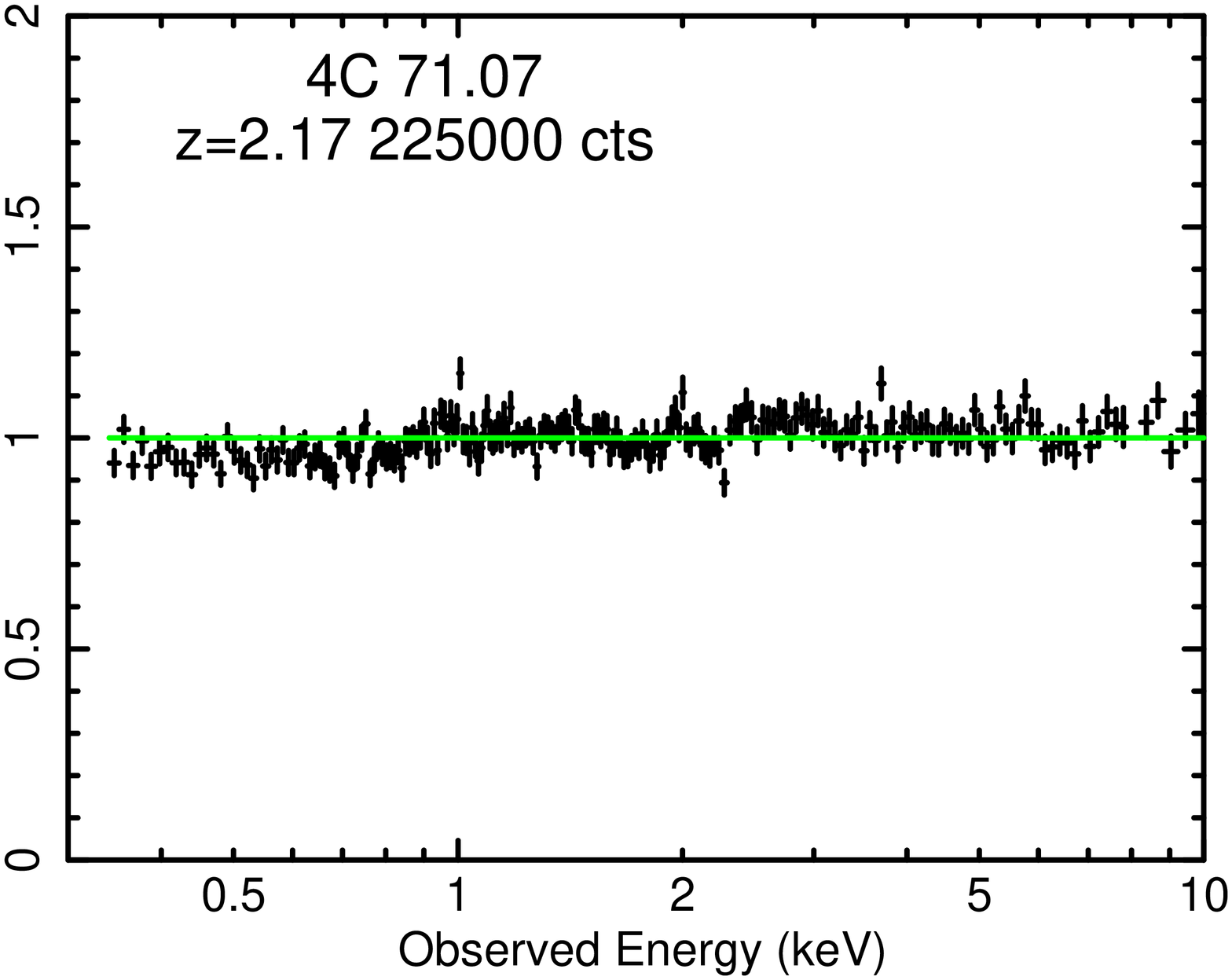}}
\caption{Data to model ratio plots for the QSO sample of Table~\ref{tab:RLQ}. 
Data are binned to conveniently represent the extra-galactic transmission functions.
Note the overall similar absorption effect, but the lack of absorption for $z < 2.5$.
Multiple spectra for a given source represent separate \xmm\ observations.}
\label{fig:RLQ} 
\end{center}
\end{figure}

The simple model of an absorbed (Galactic and host) power law fits most of the sources well as can be seen by the reduced $\chi ^2$/dof values quoted in Table~\ref{tab:RLQ}, and which are mostly close to unity.
Three quasars in our sample, however, did not yield a satisfactory fit, namely QSO\,B0014+810, QSO\,B0438-43, and PKS\,1830-210. 
The first two suffer from low count statistics at high energies, but are fitted well up to $E < 3$~keV.
C-statistic fits \citep{cash79} of the entire spectra of these two sources are good and yield 
spectral parameters that are well within the errors of the limited band $\chi ^2$ fitting. 
We could not satisfactorily fit the complex spectrum of the gravitationally lensed quasar PKS\,1830-210 \citep[][]{zhang08} with the same simple model, not even when the high energy region was ignored.
Henceforth, we refer to our sample as the 12 targets in Table~\ref{tab:RLQ}, excluding PKS\,1830-210. 
The sense of the absorption effect in this sample can be obtained from Fig.~\ref{fig:RLQ}.
In this figure, we plot the data to best-fit-model ratios for all targets of Table~\ref{tab:RLQ} except PKS\,1830-210, but after the excess column has been removed from the model, i.e. $N_H(z)$ is set to zero.
When a source was observed more than once, we plot each spectrum separately (in color in the electronic version).
The absorption effect can be seen to be consistent between different observations.

For the most part, the optical depth toward quasars is well within the scatter of the GRB optical depths (Fig.~\ref{fig:tau}) and consistent with its mean trend (Fig.~\ref{fig:igm}).
However, the four lowest redshift targets with $2 < z < 2.5$ do not show significant absorption.
Three out of these four, essentially, have upper limits of $\tau (0.5 \mathrm{keV}) \leq 0.1$ (Table~\ref{tab:RLQ}).
Formally, the quasar absorption effect is relatively uniform with an error-weighted mean of $\langle \tau (0.5 \mathrm{keV})\rangle _{z>2}\, = 0.19 \pm 0.01$, $0.35 \pm 0.09$ with no weights, and $0.32 \pm 0.08$ if the one upper limit is also included, all of which are lower than the respective GRB values (Sec.~\ref{GRB}).
The error-weighted mean optical depth of the eight quasars with $z > 2.5$ is  $\langle \tau (0.5 \mathrm{keV})\rangle _{z>2.5}\, = 0.41 \pm 0.02$, which is consistent with the GRB value, and $0.43 \pm 0.10$ with no weights.
RBS\,315 ($z = 2.69$) has somewhat high absorption \citep[see also][]{piconcelli05} with respect to the quasars with $\tau (0.5 \mathrm{keV}) = 1.08 \pm\ 0.06$, although this too is still well within the GRB scatter.
It has been suggested that perhaps the spectral curvature in RBS\,315 is intrinsic to the source and not due to absorption \citep{tavecchio07}. However, no definitive conclusion could be drawn.

The emergence of absorption at $z > 2.5$ raises an interesting connection with the re-ionization of He\,II in the IGM observed from UV spectroscopy to occur at $z = 2.7 \pm\ 0.2$ \citep{shull10}.
The contribution of He to the observed opacity at 0.3 -- 1 keV (1 -- 4 keV in the rest frame for $z = 2 - 3$ targets) is small (Fig.~\ref{fig:cs}), while the contribution of heavier elements such as C and O is appreciable.
The ionization energy of He\,II is 54.4~eV, which is comparable to that of L-shell C and O charge states. For example, the ionization energy of O\,III is 54.9~eV, and that of C\,III is 47.9~eV.
Thus, these elements are expected to be ionized into their L-shell when He\,II is ionized.
The L-shell ionization of C and O results in their K-edge (responsible for the \x\ absorption) being pushed to higher energies, and in a slightly reduced photo-ionization cross section in the sub-keV regime.
This was demonstrated recently for N in Fig.~8 of \citet{garcia09}.
The IGM \x\ opacity due to these metals will decrease more drastically once they become ionized down to their H-like state. 
This occurs for C and O at much higher ionization energies of 392.1~eV and 739.3~eV, respectively.
If the reduced absorption observed along the lines of sight to the quasars at $ 2 < z < 2.5$ is due to the intervening material being highly ionized, it would conflict with ubiquitous IGM absorption, which is expected to nearly saturate by $z \approx 2.5$ (see Sec.~\ref{IGM} and Fig.~\ref{fig:igm}).
Admittedly, the current quasar sample is small.
A larger sample is needed to conclusively determine whether indeed \x\ absorption of high-$z$ quasars increases significantly around $z = 2.5$.

\subsection{Radio quiet quasars}
\label{RQQ}

Although the above sample was selected based on the availability of high S/N \x\ spectra in the \xmm\ archive, the radio brightness of these bright \x\ sources creates a strong bias in the sample towards extremely radio loud quasars (RLQs), which are only a small fraction of the total quasar population.
Are RQQs absorbed in the soft \x s as much as the RLQs and GRBs?
The answer to this question can not benefit from the high S/N available for the RLQs.
Nevertheless, it is of considerable importance for assessing IGM absorption.
Comprehensive studies of high-$z$ RQQs have been carried out by \citet{shemmer05,shemmer06,shemmer08}, who find 
the observed soft \x\ spectrum of RQQs in these studies to be essentially featureless.
In other words, significant absorption (or soft excess) is not detected.
We obtained from the \xmm\ archive the highest S/N spectra for RQQs.
The data reduction and handling is similar to that of the RLQs in Sec.~\ref{quasars}.
We find similar results to those of \citet{shemmer05,shemmer06,shemmer08} and include these four data points in Fig.~\ref{fig:igm}.

\begin{figure}[pt!]
\begin{center}
\vglue0.0cm
{\includegraphics[angle=-90,width=8cm]{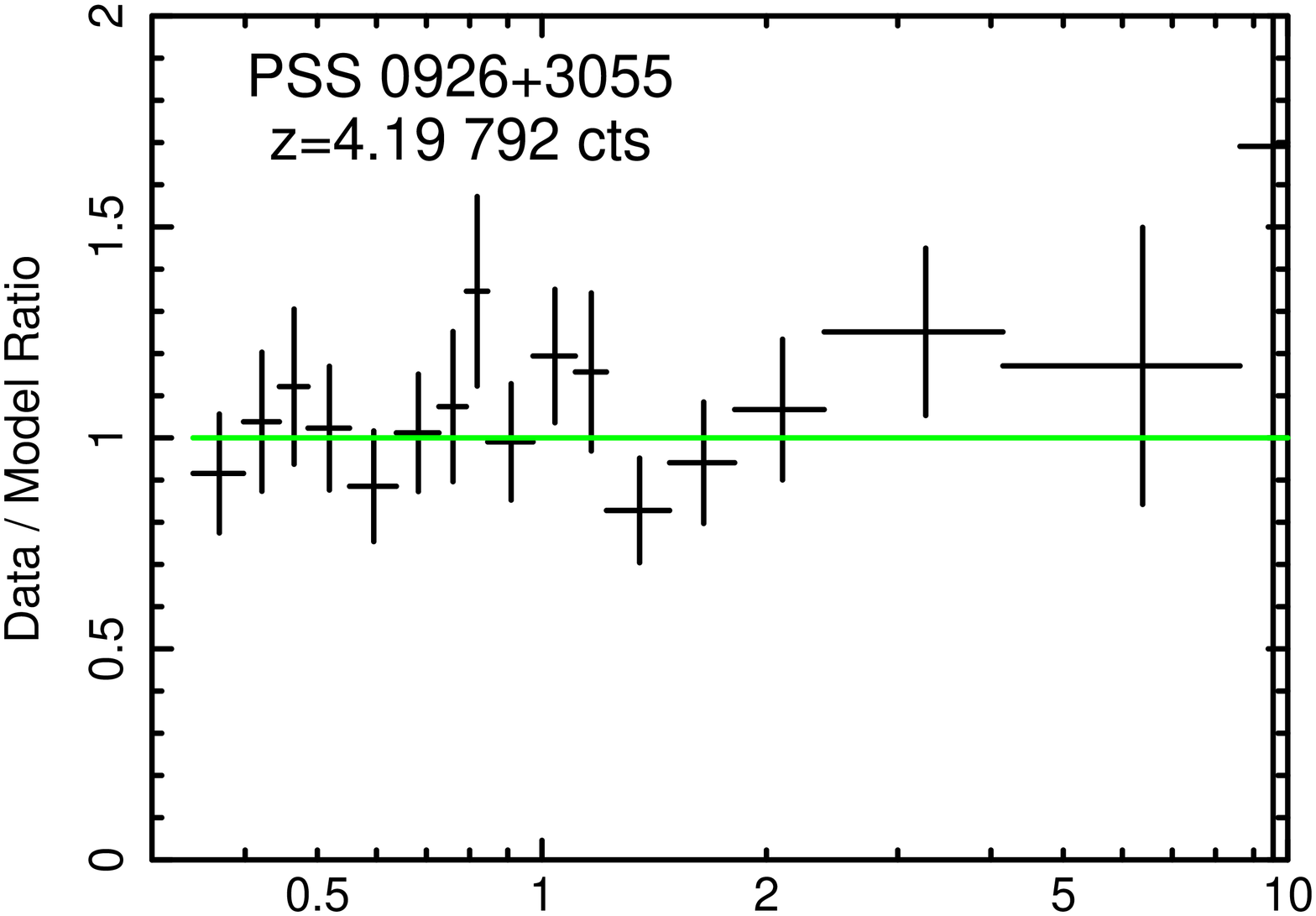}}
{\includegraphics[angle=-90,width=7.5cm]{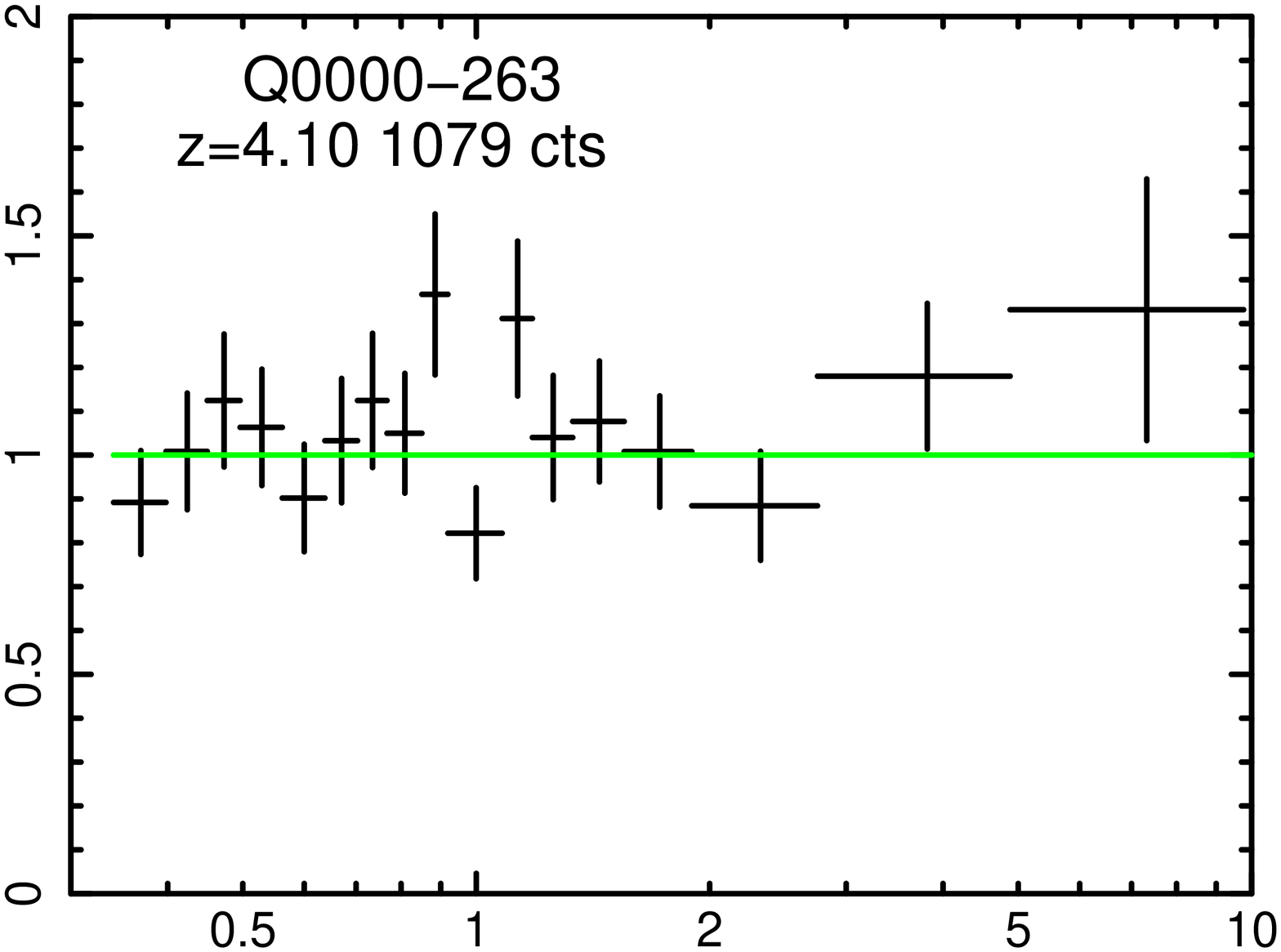}}
{\includegraphics[angle=-90,width=8cm]{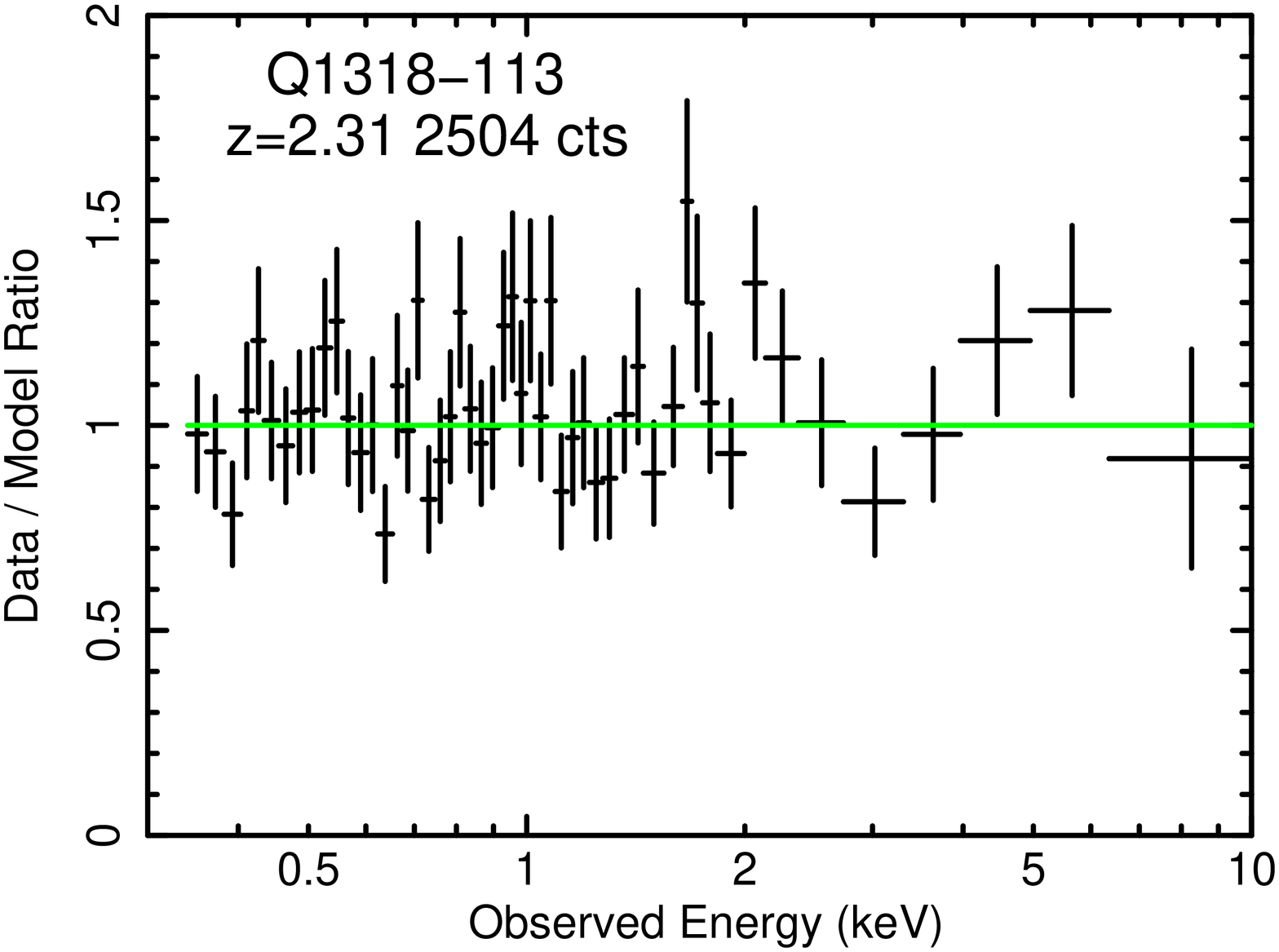}}
{\includegraphics[angle=-90,width=7.5cm]{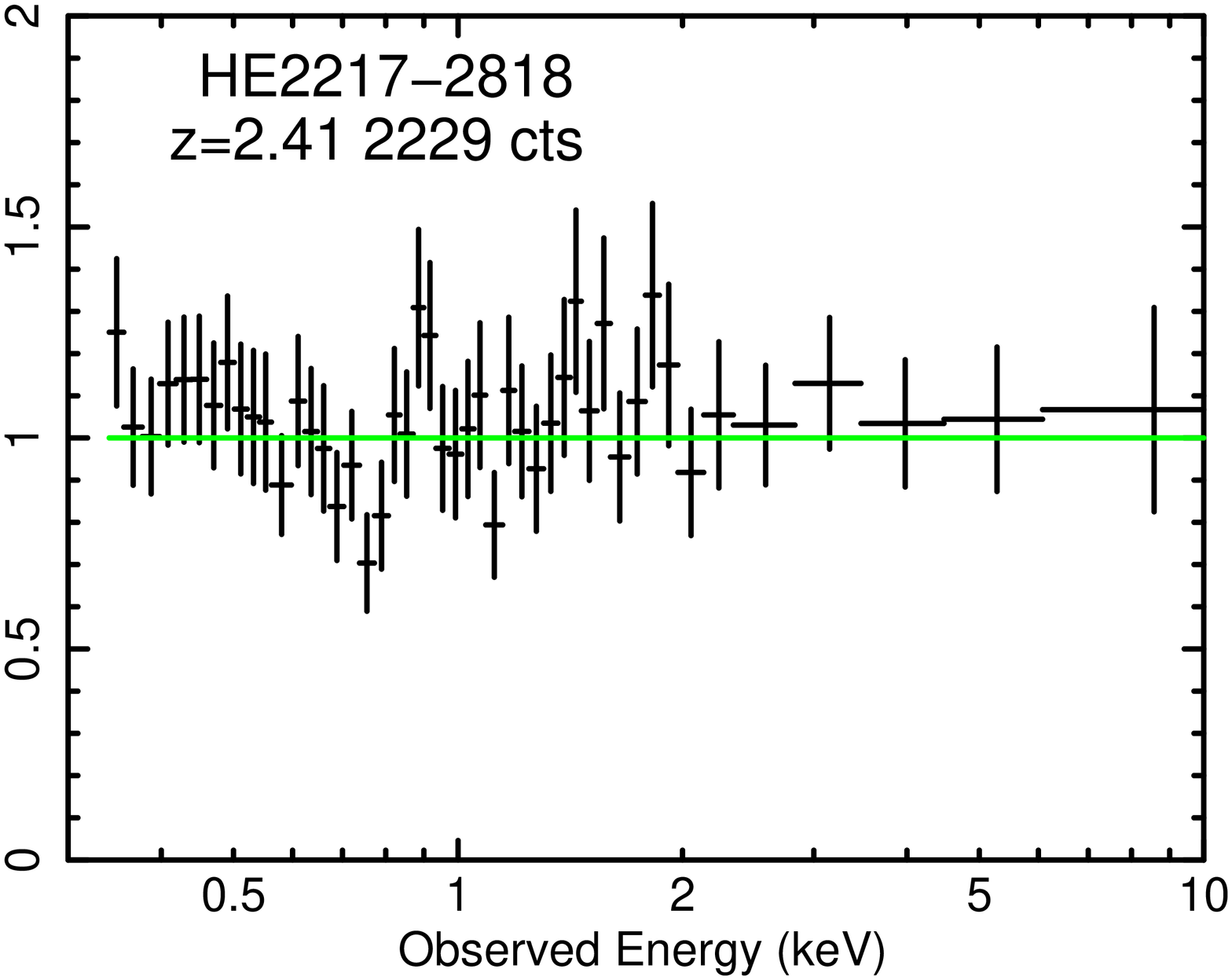}}
\caption{Data to model ratio plots for the highest-count RQQs listed in Table~\ref{tab:RQQ}. 
Plotted are in effect the extra-galactic transmission functions.
Data are binned to conveniently represent the extra-galactic transmission functions.
The lack of absorption seen for the two $z < 2.5$ sources is consistent with the RLQ results (Fig.~\ref{fig:RLQ}), while for $z > 4$ it is not. This result, however, needs to be taken with caution due to the low S/N RQQ spectra.}
\label{fig:RQQ} 
\end{center}
\end{figure}

In Fig.~\ref{fig:RQQ} we show the 4 RQQs with the most \pn\ counts.
The parameters of these sources are given in Table~\ref{tab:RQQ} and plotted in Fig.~\ref{fig:igm}.
It can be seen that the RQQ spectra are much noisier than those of the RLQs and that none of the sources reveals significant absorption.
The two RQQs with the most counts (bottom of Fig.~\ref{fig:RQQ}) are also at $z < 2.5$, where the RLQs do not show any significant absorption either (Figs.~\ref{fig:igm}, \ref{fig:RLQ}).
The two other RQQs at $z > 4$ suffer from even poorer S/N. 
Therefore, their spectra constrain the optical depth at 0.5~keV only poorly.
The mean optical depth of $\sim 0.3$ found for the RLQs is still within the errors for these two sources,
although their best fit values are lower.

Despite the limited S/N compared to the RLQs (and many GRBs), 
the lack of absorption in RQQs \citep[11 sources in][]{shemmer08} appears to be in contrast with the (diffuse IGM?) absorption effect observed in GRBs and RLQs.
RQQs in the local universe are known to have soft \x\ excess emission that can offset the absorption effect.
However, this excess that reaches $\sim 1$ keV in the quasar rest frame is not expected to affect the observed 0.5~keV opacity for sources with $z > 2$, unless its strength increases with $z$. 
To summarize this section, the sense of no absorption in RQQs is visible in Fig.~\ref{fig:RQQ}
\citep[see also the plots in][]{shemmer05,shemmer06,shemmer08},
although some of the opacity upper limits are still high.

\section{Discussion and Conclusions}
\label{concl}

The extragalactic soft \x\ opacity towards GRBs and quasars at low $z$ is dominated by absorption in the host galaxy. 
However, the contribution of a fixed host column to the optical depth $\tau$ at a given observed energy sharply drops with $z$, and the extragalactic opacity observed in the soft X-ray spectra of high-$z$ GRBs may thus have a dominant contribution from the diffuse IGM.
Due to the redshift and energy dependence of the photo-electric cross section,
the diffuse IGM opacity saturates at $z~\gsim\ 2$ and is expected to be isotropic and relatively constant  beyond that redshift.
The redshift where saturation occurs depends on the IGM metallicity.
The IGM opacity is not expected to directly correlate with the line absorption 
systems observed in the optical and UV that originate either in the host galaxy or in over-dense IGM clumps.
A large sample of \swift\ GRBs and a high S/N sample of RLQs observed with \xmm\ appear to mostly be consistent with this picture of diffuse IGM absorption.
However, a few RLQs at $z < 2.5$ and a few low S/N RQQs with very little absorption that nonetheless could represent a much larger quasar sample raise doubts.

For the IGM to produce the \x\ opacities observed towards high-$z$ GRBs and RLQs,
it needs to be relatively enriched with metals and not too highly ionized.
If indeed it is the diffuse IGM responsible for this absorption, 
it can account for a significant fraction of the currently missing baryons,
implied by big bang nucleosynthesis \citep{Steigman2007},
the observed angular power spectrum of the cosmic microwave background radiation and the Thomson opacity inferred from its polarization \citep{Komatsu2010}.
Of these baryons, in the local universe only $\sim 50\%$ are present in the galaxies, galaxy clusters and 
UV-optical IGM line systems known to date \citep[for a review see][]{bregman07}. 
Furthermore, if 90\% of the baryons are in the IGM and if the diffuse low-$z$ IGM metallicity is indeed $\sim 0.2 - 0.4$ solar as we postulate here, 
then the IGM contains the bulk of the metals in the present-day universe, compared to, say, solar metallicity in stars and galaxies that comprise only 10\% of the baryons. 

The presence of absorption in GRBs and RLQs, that both harbor powerful jets, but less in RQQs, also raises the possibility that the absorption effect has something to do with the jet.
It was discovered by \rosat\ \citep{elvis94, fiore98} that high-$z$ RLQs are much more absorbed in the \x s than RQQs, which seems to support a jet effect.
On the other hand, these authors also realized the absorption (if intrinsic) increases with $z$ and not with luminosity, which argues against a jet-physics origin. 
The fact that local RLQs generally do not have the high column densities that are found in the high-$z$ sources 
has been confirmed recently by \citet[][Table~4 therein]{galbiati05}. 
Indeed, ascribing photo-electric absorption to the jet is counter-intuitive, as the jet is not expected to comprise atomic material with bound electrons, and especially not metals, that produce the observed \x\ opacity.
Moreover, the little scatter of the observed optical depth found in this paper, and even more the tendency to an asymptotic opacity at high $z$ that require a putative intrinsic column to scale approximately with $(1+z)^{2.5}$ to offset the decreasing cross section, calls into question the realistic role jets can play in determining the soft \x\ opacity. 

Turning to the trends of intergalactic line absorption towards GRBs and quasars in the optical band does not provide a clear-cut answer to the uniqueness of absorption towards jetted sources either. 
On one hand, the number of Mg\,II intervening absorption systems towards GRBs \citep{prochter06} and blazars \citep{bergeron10} is a few times higher than that towards RQQs.
As there is no obvious reason for line sights towards GRBs to be different from those towards quasars, 
it would require metals to be entrained in the jets of GRBs and RLQs.
On the other hand, line sights towards high-$z$ RLQs seem to have a similar number density of intervening DLA systems as those towards optically-selected (again, mostly RQQs) samples \citep{ellison01}.

Prospectively, there is a clear way to distinguish a well confined absorber from a cosmologically diffuse one based on line absorption.
The detection of absorption lines, or lack thereof at high spectral resolution (and high S/N) would provide a definitive characterization of the absorber.
The CCD \x\ detectors used in this work do not have the spectral resolution required to discern lines, while the grating spectrometers on board \xmm\ do not have the sufficient effective area to provide adequate spectra, at least not with the modest exposures available to date.
High S/N high-resolution spectra of the quasars either through a long \xmm\ exposure, or with future instruments should be able to unambiguously determine whether \x\ absorption of high-$z$ sources is due to their hosts or due to the diffuse IGM.

\acknowledgments
The authors thank A. Heller for assistance with the GRB data extraction, and S. Kaspi and A. Nusser for early discussions.
We acknowledge important insights by D. Chelouche on Mg\,II absorption systems, and by O. Shemmer on \x\ absorption of RQQs.
E. B. is grateful to R. Mushotzky for insightful discussions on He\,II re-ionization and for the hospitality at the U of MD where part of this work was carried out.
This research was supported by an ISF grant and makes use of data supplied by the UK Swift Science Data Centre at the University of Leicester.

\newpage
\begin{deluxetable}{llcrcccc}
\tabletypesize{ \footnotesize }
\tablecolumns{8} \tablewidth{0pt}
\tablecaption{High-$z$ Quasars}
\tablehead{
   \colhead{Source} &
   \colhead{\xmm} &
   \colhead{$z$} &
   \colhead{photons} &
   \colhead{$N_H$(Gal.)} &
   \colhead{$N_H$($z$) \tablenotemark{a}} &
   \colhead{$\tau$(0.5 keV)}\tablenotemark{b} &
   \colhead{$\chi ^2$/dof} 
\\
   \colhead{} &
  \colhead{Obs ID(s)} &
 \colhead{} &
   \colhead{} &
   \colhead{10$^{20}$cm$^{-2}$} &
   \colhead{10$^{22}$cm$^{-2}$} &
   \colhead{} &
   \colhead{}
}
\startdata
GB6\,B1428+4217 &  0111260101/701 &  4.715 & 12580 & 1.18 & 2.1 $\pm$ 0.5  & 0.24 $\pm$ 0.06 & 1.04\\ 
& 0212480701 &  &  & &  &  &  \\
PMN\,J0525-3343 & 0050150101/301  & 4.413 & 28800 & 2.28 & 1.9 $\pm$ 0.3  & 0.26 $\pm$ 0.04 & 0.98 \\
& 0149500601/701/801/901 &  &  & &  &  &  \\
& 0149501001/101 &  &  & &  &  &  \\
RXJ\,1028.6-0844 & 0093160701, 0153290101 & 4.276 & 6250 & 4.49 & 1.8 $\pm$ 0.7  & 0.27 $\pm$ 0.10 & 0.98 \\
QSO\,B0014+810 \tablenotemark{c} & 0112620201 & 3.366 & 12500 & 13.6 & 1.8 $\pm$ 0.5  & 0.41 $\pm$ 0.11 & 0.98 \\
PKS 2126-158 & 0103060101 & 3.268 & 35200 & 4.92 & 1.6 $\pm$ 0.2  & 0.39 $\pm$ 0.05 & 0.98 \\
QSO\,B0438-43 \tablenotemark{c} & 0104860201 & 2.852 & 7400 & 1.35 & 1.8 $\pm$ 0.3  & 0.59 $\pm$ 0.12 & 1.08 \\
RBS\,315 & 0150180101 & 2.690 & 69950 & 9.26 & 2.9 $\pm$ 0.2  & 1.08 $\pm$ 0.06 & 1.00 \\
PKS\,2351-154 &0203240201 & 2.675 & 9000 & 2.51 & 0.6 $\pm$ 0.2  & 0.21 $\pm$ 0.07 & 1.05 \\
PKS\,1830-210 & 0204580201/301/401 & 2.507 & 75500 & 20.2 & 18 $\pm$ 0.6  & 7.1 $\pm$ 0.2 & 1.18 \\
QSO\,J0555+3948 & 0300630101 & 2.363 & 5000 & 28.2 & 0.49 $\pm$ 0.44  & 0.22 $\pm$ 0.20 & 0.97 \\
PKS\,2149-306 & 0103060401  & 2.345 & 36200 & 1.61 & 0.08 $\pm$ 0.07  & 0.036 $\pm$ 0.032 & 1.00 \\
PKS\,0237-230 &  0300630301 & 2.225 & 12550 & 2.16& 0.10 $\pm$ 0.10 & 0.05 $\pm$ 0.05  & 1.06 \\
4C\,71.07 &  0112620101 & 2.172 & 225450 & 2.85 & 0.09 $\pm$ 0.03  & 0.05 $\pm$ 0.02 & 1.08\\
\tablenotetext{a}{~Measured assuming an absorber at the host at redshift $z$.}
\tablenotetext{b}{~Deduced from $N_H (z)$ using Eq.~\ref{tau*}.}
\tablenotetext{c}{~Fitted only up to 3~keV.}
\enddata
\label{tab:RLQ}
\end{deluxetable}

\newpage
\begin{deluxetable}{llcrcccc}
\tabletypesize{ \footnotesize }
\tablecolumns{8} \tablewidth{0pt}
\tablecaption{Radio Quiet Quasars}
\tablehead{
   \colhead{Source} &
   \colhead{\xmm} &
   \colhead{$z$} &
   \colhead{photons} &
   \colhead{$N_H$(Gal.)} &
   \colhead{$N_H$($z$) \tablenotemark{a}} &
   \colhead{$\tau$(0.5 keV)} \tablenotemark{b} &
   \colhead{$\chi ^2$/dof} 
\\
   \colhead{} &
  \colhead{Obs ID(s)} &
 \colhead{} &
   \colhead{} &
   \colhead{10$^{20}$cm$^{-2}$} &
   \colhead{10$^{22}$cm$^{-2}$} &
   \colhead{} &
   \colhead{}
}
\startdata
PSS\,0926+3055 & 0200730101 & 4.190 & 800 & 1.89 & 0.35$^{+1.5}_{-0.35}$ & 0.06$^{+0.24}_{-0.06}$ & 1.02 \\
Q\,0000-263 & 0103060301 & 4.111 & 1100 & 1.67 & 0.5$^{+1.3}_{-0.5}$  & 0.08$^{+0.21}_{-0.08}$ & 1.26 \\
HE\,2217-2818 & 0302380401 & 2.414 & 2250 & 1.28 & 0$^{+0.1}_{-0}$  & 0$^{+0.05}_{-0}$ & 0.90 \\
Q\,1318-113 & 0402070301 & 2.306 & 2500 & 2.22 & 0.1$^{+0.26}_{-0.1}$  & 0.05$^{+0.12}_{-0.05}$ & 1.07 \\
\tablenotetext{a}{~Measured assuming an absorber at the host at redshift $z$}
\tablenotetext{b}{~Deduced from $N_H (z)$ using Eq.~\ref{tau*}}
\enddata
\label{tab:RQQ}
\end{deluxetable}

\newpage

\newpage

\newpage

\newpage

\newpage

\newpage

\end{document}